\documentclass[11pt,draftcls,onecolumn] {IEEEtran}
\usepackage{bbm}
\usepackage{psfrag}
\usepackage{color}
\usepackage[compress]{cite}
\usepackage[pdftex]{graphicx}
\usepackage{amssymb}
\usepackage[cmex10]{amsmath}
\usepackage{amsthm}
\usepackage{algorithm}
\usepackage{algorithmic}
\usepackage{array,enumerate}
\usepackage{multirow}
\usepackage[tight, footnotesize]{subfigure}
\usepackage{url}
\usepackage{xparse}
\graphicspath{{./Figures/}}

\newtheorem{definition}{Definition}

\newtheorem{theorem}{Theorem}

\newtheorem{lemma}{Lemma}
\theoremstyle{definition} 

\newcommand{\dt}{\bar{d}(t)}

\newcommand{\floor}[1]{{\left\lfloor {#1} \right\rfloor}}
\newcommand{\ceil}[1]{{\lceil {#1} \rceil}}

\newcommand{\norm}[1]{\left\|#1\right\|}

\DeclareDocumentCommand \GI { o } {%
  \IfNoValueTF {#1} {%
    G_{\textnormal{I}}%
  }{%
    G_{\textnormal{I}, #1}%
  }%
}

\newcommand{\tobs}{t_{\textnormal{obs}}}
\newcommand{\nobs}{n_{\textnormal{obs}}}
\newcommand{\leaf}[1]{l_{{#1}}}
\newcommand{\tleaf}[1]{\tilde{l}_{{#1}}}
\newcommand{\ttree}[1]{\tilde{T}_{#1}}
\newcommand{\VI}{V_{\textnormal{I}}}
\newcommand{\sss}{v^*}
\newcommand{\est}{\hat{v}}

\newcommand{\dpath}{P_d}
\newcommand{\Vsp}{V_{\textnormal{sp}}}
\newcommand{\admincost}{c_{a}}
\newcommand{\admingain}{g_{a}}
\newcommand{\admind}{d_{a}}
\newcommand{\sourcegain}{g_{s}}
\newcommand{\sourcecost}{c_{s}}
\newcommand{\sourced}{d_{s}}
\newcommand{\MIS}[1]{{\Lambda_{#1}}}
\newcommand{\MISSET}[1]{{\mathcal{M}(#1)}}
\newcommand{\JC}{\textnormal{JC}}

\DeclareDocumentCommand \DIS { o } {%
  \IfNoValueTF {#1} {%
    \textnormal{DIS}%
  }{%
    \textnormal{DIS}(#1)%
  }%
}

\usepackage[normalem]{ulem}

\begin{document}
\title{Infection Spreading and Source Identification: A Hide and Seek Game}

\author{Wuqiong~Luo,~\IEEEmembership{Student Member,~IEEE}, Wee~Peng~Tay,~\IEEEmembership{Senior Member,~IEEE} and Mei~Leng,~\IEEEmembership{Member,~IEEE}
\thanks{An abridged version of this paper has been submitted to 2015 IEEE/ACM International Conference on Advances in Social Networks Analysis and Mining. This work was supported in part by the Singapore Ministry of Education Academic Research Fund Tier 2 grants MOE2013-T2-2-006 and MOE2014-T2-1-028. W. Luo and W.~P. Tay are with the Nanyang Technological University, Singapore. M. Leng is with the Temasek Laboratories@NTU, Singapore. E-mail: \texttt{\{wqluo, wptay, lengmei\}@ntu.edu.sg}.
}}
\maketitle

\begin{abstract}
The goal of an infection source node (e.g., a rumor or computer virus source) in a network is to spread its infection to as many nodes as possible, while remaining hidden from the network administrator. On the other hand, the network administrator aims to identify the source node based on knowledge of which nodes have been infected. We model the infection spreading and source identification problem as a strategic game, where the infection source and the network administrator are the two players. As the Jordan center estimator is a minimax source estimator that has been shown to be robust in recent works,  we assume that the network administrator utilizes a source estimation strategy that can probe any nodes within a given radius of the Jordan center. Given any estimation strategy, we design a best-response infection strategy for the source. Given any infection strategy, we design a best-response estimation strategy for the network administrator. We derive conditions under which a Nash equilibrium of the strategic game exists. Simulations in both synthetic and real-world networks demonstrate that our proposed infection strategy infects more nodes while maintaining the same safety margin between the true source node and the Jordan center source estimator.
\end{abstract}


\section{Introduction}\label{sec:Introduction}
With the increasing popularity of online social networks like Facebook, Twitter and Google+ \cite{Viswanath2009,Kumar2010,Cha2010,Gundotra2013}, more and more people are getting news and information via social networks instead of traditional media outlets. According to a study from the Pew Research Center, about 30\% of Americans now get news from Facebook \cite{PewResearchCenter2014}. Due to the interactive nature of online social networks, instead of passively consuming news, half of social networks users actively share or repost news stories, images or video, and 46\% of them discuss news issues within their social circles \cite{PewResearchCenter2014}. As a result, a piece of information or a rumor posted by a social network user can be reposted by other users and spread quickly on the underlying social network and reach a large number of users in a short period of time \cite{Han2008, Weng2010, Bakshy2011}. We say that such users or nodes in the network are ``infected''. A widely spread rumor or misinformation can lead to reputation damage \cite{RumorJackieChan2013}, political consequences \cite{Garrett2011}, and economic damage \cite{RumorWhiteHouse2013}. The network administrator may want to identify the rumor source in order to catch the culprit, control the damage, and counter the rumor influence. Here, the term ``network administrator'' is used in a very broad sense to include anyone (e.g., regulatory authorities and researchers) who has been given access to data about the network topology and infected nodes.

Another example of an infection spreading is that of a malicious node in a computer network whose goal is to spread a virus throughout the network. The virus can be a spam bot that is not easily detected \cite{Dittrich2008} (e.g., when the Mariposa botnet was dismantled in 2009, it had infected over 8 million computers \cite{Thompson2009}), and the network administrator is alerted to the virus infection only at a much later time. Motivated by these applications, many recent research works \cite{Shah2011, Luo2013, Dong2013, Zhu2012, Luo2014, Luo2014arxiv} have focused on the problem of identifying rumor or infection sources in a network under various spreading models. In all these works, the source is assumed to be ``dumb'', and whether a susceptible node becomes infected or not follows a stochastic process that is not controlled by the source. Under this simplified assumption, the works \cite{Shah2011, Luo2013, Dong2013, Zhu2012, Luo2014, Luo2014arxiv} show that source estimators can be constructed so that the true source can be identified with high probability to within a fixed number of hops.

In many applications, the source may wish to maintain anonymity while spreading the infection to as many users as possible. An example is the now defunct anonymous social networking app Secret \cite{Secret}, which allowed smart phone users to share information and repost a posting anonymously among his device contacts or Facebook friends. In February 2014, Secret was used to spread the false rumor that Evernote Corporation was going to be acquired, which prompted the CEO to subsequently issued a public denial \cite{RumorEvernote2014}. %
Messaging services including Wickr \cite{Wickr} and FireChat \cite{FireChat} have been used in civil protests like those in Hong Kong in 2014 \cite{Wiki_HKprotests,CNN_FireChat}. Government authorities may trace the initiators of certain protest events even if the messages are encrypted through the use of source identification algorithms that do not rely on message contents or metadata \cite{Shah2011, Luo2013, Dong2013, Zhu2012, Luo2014}. Therefore, in distributing information, civil protest leaders may design an infection strategy that carefully controls the rate of information spreading in order to obfuscate their identities. In the example of spam bot infection spreading, the perpetrator also wants an infection strategy that controls the rate of the virus spreading to avoid being caught by the authorities while spreading the virus to as many computers as possible.
The recent work \cite{Fanti2014} introduces a messaging protocol, which guarantees obfuscation of the source under the assumption that the network administrator utilizes the maximum likelihood (ML) estimator to identify the source, and when the underlying network is an infinite regular tree. Moreover, simulations are provided in \cite{Fanti2014} to verify the performance of the messaging protocol on irregular trees and general networks.

With prior knowledge that the infection source may try to avoid detection, the network administrator needs to adapt its estimation strategy to increase its chance of identifying the infection source. On the other hand, if the source has prior knowledge of the estimation strategy, it needs to further adapt its own infection strategy, and so on. The source and network administrator is thus playing a ``hide and seek'' game of infection spreading and source identification. This complex dynamic can be modeled as a strategic game with the source and network administrator as the two players of the game. To the best of our knowledge, studying infection spreading and source identification as a strategic game is novel since previous works like \cite{Shah2011, Luo2013, Dong2013, Zhu2012, Luo2014, Luo2014arxiv} focus only on the estimation strategy, while the work \cite{Fanti2014} focuses only on the infection strategy.

In this paper, we study best-response strategies for \emph{both} the source and network administrator for trees from a game theoretic perspective, whereas \cite{Fanti2014} develops an \emph{order-optimal} infection strategy for the source for infinite regular trees (which are special cases of expanding trees), and extends heuristically to more general networks. In \cite{Fanti2014}, the network administrator is assumed to adopt the ML estimation strategy, whereas we assume that the network administrator is allowed to tune a Jordan center based estimation strategy (see Section \ref{subsec:problem_formulaiotn:network_administrator} for justifications of our estimation strategy choice). %
In our current work, we assume that the network administrator becomes aware of the infection only when the number of infected nodes exceeds a given threshold, and only then it makes an observation of the infection status of the nodes. (The problem becomes trivial if the network administrator is constantly monitoring all the nodes in the network.) For example, a perpetrator who aims to manipulate the stock price of a company may start to spread a false rumor about the company on a social network. The regulatory authority do not have enough resources to monitor the whole network all the time and for all possible events. Therefore, it becomes aware of the false rumor only when the number of infected nodes becomes sufficiently large.
Our work can also be applied to the case where the source has an estimate of when the network administrator discovers the infection. This can be the case in the previous stock price manipulation example when the perpetrator first spreads a rumor within his social network using private messages to collude with other users, and then all colluding users post the rumor publicly on the stock's initial public offering day in order to manipulate its price and profit from it.

Our main contributions are the following:
\begin{enumerate}[(i)]
	\item\label{cont:game} We formulate a strategic game in which the network administrator and infection source are the players. The network administrator uses a source estimator in which it can probe any nodes within a given radius of a randomly chosen Jordan center of the observed infection graph. A larger probe or estimation radius ensures that the source can be identified but incurs a higher cost. %
The infection source uses an infection strategy in which the rate of infection over each edge in the network can be controlled in order to achieve a minimum safety margin to the Jordan centers. The source is rewarded for each infected node, and penalized if it is identified by the network administrator.
	\item Given a safety margin for the infection source, we show that the best-response strategy for the network administrator is to use the Jordan centers as the source estimator or adopt an estimation radius equal to the safety margin. We derive conditions under which each of these strategies are optimal.
	\item Given an estimation radius for the network administrator, we show that the optimal safety margin for the infection source when the underlying network is a tree, is either zero or one more than the estimation radius. We derive an infection strategy, called the Dominant Infection Strategy (DIS), which maximizes the number of infected nodes subject to a given safety margin.
	\item We derive conditions under which a Nash equilibrium for the strategic game in \eqref{cont:game} exists. We show that when a Nash equilibrium exists, the best response for the network administrator is to adopt the Jordan center estimator. This gives a game-theoretic interpretation to the Jordan center estimator, in addition to being a universally robust estimator (which we showed previously in \cite{Luo2014arxiv}).
\end{enumerate}

Our problem of finding the best-response infection strategy is related to the influence maximization problem, which aims to find a subset of influential nodes to maximize the expected number of nodes that are ``influenced'' or infected by the chosen subset \cite{Domingos2001, Richardson2002}, and is shown by \cite{Kempe2003} to be a NP-hard optimization problem. Approximate solutions have been extensively investigated by various researchers \cite{Leskovec2007_2,Chen2009,Chen2010}. The main difference between our work and the influence maximization problem is that the source node in our problem is fixed, and we seek an infection strategy, given any source node, to maximize the set of infected nodes, subject to a safety margin to the Jordan center of the infection graph.

The rest of this paper is organized as follows. In Section \ref{sec:problem_formulation}, we present the system model, assumptions and provide a game-theoretic problem formulation. In Section \ref{sec:optimal_estimation_strategy}, we show the best-response estimation strategy for the network administrator given any infection strategy. In Section \ref{sec:optimal_infection_strategy_trees}, we propose a best-response infection strategy for the source given any estimation strategy. In Section \ref{sec:nash_equilibrium}, we derive conditions under which a Nash equilibrium of the strategic game exists. We present simulation results in Section \ref{sec:simulation_results} to evaluate the effectiveness of the proposed strategies on various synthetic and real networks. Finally we conclude and summarize in Section \ref{sec:conclusion}.

\section{Problem Formulation}\label{sec:problem_formulation}
In this section, we first describe our system model and assumptions, and then we provide a game-theoretic problem formulation for the infection spreading and source identification.

Consider an undirected graph $G(V,E)$ representing a social network, where $V$ is the set of vertices or nodes, and $E$ is the set of edges. Because of technical difficulties, our analysis and strategy design assume that $G$ is a tree, as is commonly done in the literature \cite{Shah2011, Luo2013, Dong2013, Zhu2012, Luo2014, Fanti2014}. We will however apply our strategies heuristically to general networks in our simulations in Section \ref{sec:simulation_results}.

We assume that there is a single source node $v^*\in V$ at time $0$. An infection can pass from one node to another. For example in the case of an online social network, a user may post a rumor he sees in the posting of his friend using his own account. An infected node remains infected throughout, and has the capability of infecting its neighbors at a deterministic rate. For any edge $(i,j) \in E$, we let $\mu(i,j)$ to be the time it takes for an infected node $i$ to infect its susceptible neighbor $j$, which we call the infection time associated with the edge $(i,j)$. Let $\lambda(i,j) = 1/\mu(i,j)$ be the infection rate of $(i,j)$. For any pair of nodes $v$ and $u$ in $G$, let $d(v,u)$ be the number of hops in the shortest path between $v$ and $u$, which is also called the \emph{distance} between $v$ and $u$. %
For any edge $(i,j)$ with $d(v^*,i)=m$ and $d(v^*,j)=m+1$, we assume $\lambda(i,j)$ is uniformly upper bounded by a maximum infection rate $\bar{\lambda}_m > 0$. In examples like rumor spreading, $\bar\lambda_m$ is non-increasing in $m$ as it becomes more difficult for an infected node further away from the source to infect another susceptible node.
We assume that the network administrator observes one snapshot of all the infected nodes in the network, and tries to estimate the source at the time $\tobs$ when the number of infected nodes first exceeds a threshold $\nobs>1$. We call $\tobs$ the observation time, and $\nobs$ the observation threshold.

For any time $t > 0$, let
\begin{align}\label{def:bard}
\bar{d}(t) = \max \left\{k : \sum_{m=0}^{k-1}\bar{\lambda}_m^{-1} \le t \right\}.
\end{align}
Since the infection rate of each edge is upper bounded by its respective $\bar{\lambda}_m$, the maximum number of hops the infection can spread from the source in time $t$ is $\dt$. We assume that the graph is sufficiently large so that $\bar{d}(\tobs)\le \bar{d}(v^*, V)$, i.e., the network administrator observes the infection graph before the infection can spread to all nodes in the network.

For any pair of nodes $v$ and $u$ in $G$, let $\rho(v, u)$ to be the shortest path from $v$ to $u$, and the infection time of $\rho(v,u)$ to be
\begin{align}
\mu(v,u) &= \sum_{(i,j)\in \rho(v,u)}\mu(i,j) \nonumber \\
&= \sum_{(i,j)\in \rho(v,u)}\frac{1}{\lambda(i,j)}. \label{eqn:weight_of_path}
\end{align}
The collection of infection rates $\Lambda = \{\lambda(i,j) : (i,j) \in E\}$ is called an infection strategy for the source node. Given any infection strategy $\Lambda$, we denote the set of infected nodes at time $\tobs$ as
\begin{align}
\VI = \{ u \in G: \mu(\sss,u) \le \tobs \}. \label{eqn:infected_nodes}
\end{align}
We sometimes use $\VI(\Lambda)$ instead of $\VI$ to indicate that the given set of infected nodes resulted from the infection strategy $\Lambda$.
Let $\GI$ to be the minimum connected subgraph of $G$ that spans $\VI$, which we call the \emph{infection graph} at time $\tobs$.

Throughout this paper, we let $|X|$ denote the \emph{expected} number of nodes in the random set $X$ conditioned on the infection graph, and $\mathbf{1}_{A}$ denote an indicator function with value 1 iff the clause $A$ is true.

\subsection{Network Administrator} \label{subsec:problem_formulaiotn:network_administrator}
At the observation time $\tobs$, the network administrator observes the infection graph $\GI$, and tries to estimate the infection source. Although $\nobs$ is known to the network administrator, since it does not know the starting time that the source begins its infection spreading, it does not know the amount of elapsed time $\tobs$. We assume that the network administrator can choose a subset of nodes to investigate, which we call the suspect set. It is important for the network administrator to decide which subset of nodes to investigate in order to minimize the cost and maximize its chance of identifying the infection source. %
In the same spirit as \cite{Shah2011, Luo2013, Dong2013, Zhu2012, Luo2014}, the network administrator is assumed to have limited knowledge of the underlying infection spreading process, and its estimation strategy can only depend on the observed infection graph $\GI$. In the following, we present the definition of the Jordan center and then introduce a class of estimation strategies based on the Jordan center.

Given any set $A \subset V$, denote the largest distance between $v$ and any node $u \in A$ to be
\begin{align*}
    \bar{d}(v,A)=\max_{u \in A} d(v,u).
\end{align*}
For any infection strategy $\Lambda$, we call the largest distance $\bar{d}(v,\VI)$ between $v$ and any infected node the \emph{infection range} of $v$. We let
\begin{align}
\text{JC} = \{v: \bar{d}(v, \VI) = \min_{u \in G}\bar{d}(u, \VI)\}, \label{eqn:Jordan_centers}
\end{align}
to be the set of nodes with minimum infection range, which are known as the \emph{Jordan centers} of $\GI$ \cite{Wasserman1994}. It is shown in \cite{Zhu2012} that if $G$ is a tree, then $|\text{JC}| \leq 2$.

When no prior knowledge of the infection source is available, any node in $\VI$ is equally likely to be the source as infection rates over different edges can be heterogeneous. Therefore, a Jordan center is a minimax source estimator that minimizes the largest distance to any infected node.  It has also been shown in \cite{Zhu2012,Luo2014, Luo2014arxiv} that the Jordan center is a robust source estimate. Another popular estimator is the ML estimator. However, \cite{Fanti2014} shows that it is possible to design an infection strategy so that the probability of the ML estimator being the true source is approximately $1/|\VI|$, i.e., all the infected nodes are considered by the network administrator to be approximately equally likely to be the source. If $\tobs$ is large, then the ML estimator performs badly. The Jordan center estimator does not have this problem since there are at most two Jordan centers in any tree. As such, we assume that the network administrator chooses the suspect set $\Vsp(\admind)$ to be the set of infected nodes within $\admind\geq 0$ hops from an arbitrarily chosen Jordan center $u$, i.e.,
\begin{align}
\Vsp(\admind) & = \{v\in \VI: d(v, u) \le \admind \}. \label{eqn:admin_investigate_subset}
\end{align}
We call $\admind$ the \emph{estimation radius}. Note that $\Vsp(\admind)$ depends only on the observed infection graph $\VI$. The strategy of the network administrator is denoted using $\admind$.

We let $\sourced$ to be the distance between the actual source and the Jordan centers, i.e.,
\begin{align}
\sourced(\Lambda) = \min_{u \in \JC} d(\sss, u).  \label{eqn:source_safety_margin}
\end{align}
We call $\sourced(\Lambda)$ the \emph{safety margin} of the source achieved by the infection strategy $\Lambda$.

If $\admind \ge \sourced(\Lambda)$, then the infection source is in the suspect set, and the network administrator has a non-negative probability of identifying the source. The network administrator obtains an expected gain $\admingain(d_a,\VI)$, which we assume to be non-increasing in $d_a$. For example, if the network administrator only has access to the infection graph and has no additional prior information, its best strategy is to uniformly choose a node from the set of suspects as the estimated source node. Its expected reward is then inversely proportional to the number of nodes in $\Vsp(d_a)$. In another application, the network administrator may have side information that allows it to always correctly identify the source node if it is included in the suspect set. In this case, we let the expected reward to be $\admingain(d_a,\VI)=\admingain$. Some positive cost $\admincost(\Vsp(\admind))$ is incurred for probing nodes in $\Vsp(\admind)$. We assume that $\admincost(\Vsp(\admind))$ is an increasing function of the estimation radius $d_a$. We let the utility function of the network administrator be
\begin{align}
u_{a}(\admind,\Lambda) &=  -\admincost(\Vsp(\admind)) + \admingain(d_a, \VI) \mathbf{1}_{\admind \ge \sourced(\Lambda)}. \label{eqn:admin_utility}
\end{align}
The network administrator's utility function depends on $\Lambda$ only through its infection graph $\VI$, which the administrator observes at time $\tobs$, and the safety margin $d_s(\Lambda)$. Although the network administrator's utility function depends on the safety margin of the source, it does not know a priori the safety margin chosen by the source. In this paper we perform a game theoretic analysis of the estimation strategy used.

\subsection{Infection Source}

Suppose that the network administrator uses an estimation radius of $d_a\geq 0$. We assume that the observation time $\tobs$ is unknown to the infection source, but the source knows the infection threshold $\nobs$ at which the network administrator will attempt to identify it. We show in Section \ref{sec:optimal_infection_strategy_trees} that $\tobs$ can be computed from $\nobs$. For each node that is infected, the source is rewarded with a gain $\sourcegain$. A positive cost $\sourcecost(d_a)$ is incurred if it falls within the suspect set of the network administrator. We assume that $\sourcecost(d_a)$ is a non-decreasing function of the estimation radius $d_a$. The utility function of the source adopting the infection strategy $\Lambda$ is given by
\begin{align}
u_{s}(\admind, \Lambda) & = \sourcegain |\VI| - \sourcecost(d_a) \mathbf{1}_{\admind \ge \sourced(\Lambda)}.\label{eqn:source_utility}
\end{align}

\subsection{Strategic Game}
We model the infection spreading and source identification as a strategic game where the network administrator and the infection source are the two players. The utility functions of the two players are given in \eqref{eqn:admin_utility} and \eqref{eqn:source_utility}, respectively. Given any infection strategy $\Lambda$ (or more specifically, the safety margin $\sourced(\Lambda)$), the network administrator finds the \emph{best-response estimation strategy} with estimation radius $\admind^*$ that maximizes its utility function, i.e.,
\begin{align*}
\admind^* = \arg \max_{\admind} u_{a}(\admind, \Lambda).
\end{align*}
On the other hand, given any estimation radius $\admind$, the infection source finds the \emph{best-response infection strategy} $\Lambda^*$ that maximizes its utility function, i.e.,
\begin{align*}
\Lambda^* = \arg \max_{\Lambda} u_{s}(\admind, \Lambda).
\end{align*}
If there exists a pair $(\admind^*, \Lambda^*)$ such that given $\Lambda^*$, the best-response estimation strategy has estimation radius $\admind^*$; and given $\admind^*$, the best-response infection strategy is $\Lambda^*$, then $(\admind^*, \Lambda^*)$ is a Nash equilibrium of the strategic game \cite{Osborne1994}.

In Sections \ref{sec:optimal_estimation_strategy} and \ref{sec:optimal_infection_strategy_trees}, we find the best-response estimation strategy and the best-response infection strategy for the network administrator and the infection source, respectively. In Section \ref{sec:nash_equilibrium}, we derive the Nash equilibrium of the strategic game.

\section{Best-response Estimation Strategy for the Network Administrator}\label{sec:optimal_estimation_strategy}
In this section, we derive the best-response estimation strategy for the network administrator. In this paper, we assume that the network administrator utilizes the Jordan center based estimation strategy, which is characterized by the estimation radius $\admind$. Given the infection strategy $\Lambda$ with safety margin $\sourced(\Lambda) = d_s$, the network administrator chooses an optimal estimation radius to maximize its utility function.

We first consider the case where $\sourced = 0$. In this case, the inequality $\admind \ge \sourced$ always holds. From \eqref{eqn:admin_utility}, the network administrator's utility function becomes
\begin{align*}
u_{a}(\admind, \Lambda) & = -\admincost(\Vsp(\admind)) + \admingain(d_a,\VI),
\end{align*}
which is decreasing in $d_a$. Therefore, the optimal estimation radius is given by $d_a=0$.

Now suppose that $\sourced > 0$. We claim that the estimation radius of a best-response estimation strategy is either 0 or $\sourced$. To prove this claim, it suffices to show the following inequalities:
\begin{align}
u_a(0, \Lambda) > u_a(\admind', \Lambda), \ \ \ &\forall \ \admind' \in (0, d_s); \label{ineqn:ua_not_identified}\\
u_a(\sourced, \Lambda) > u_a(\admind', \Lambda), \ \ \ &\forall \ \admind' \geq \sourced+1. \label{ineqn:ua_identified}
\end{align}

We first show the inequality \eqref{ineqn:ua_not_identified}. Since $0 < \admind' <\sourced$, the gains for both the estimation strategy with $\admind=0$ and the estimation strategy with $\admind=\admind'$ are 0. The inequality \eqref{ineqn:ua_not_identified} then holds because $\admincost(\Vsp(0)) < \admincost(\Vsp(\admind'))$. We next show the inequality \eqref{ineqn:ua_identified}. The inequality $\admind \ge \sourced$ holds for both the estimation strategy with $\admind=\sourced$ and the estimation strategy with $\admind=\admind' \geq \sourced+1$. Since $\admincost(\Vsp(\sourced))< \admincost(\Vsp(\admind'))$ and $g_a(d_s,\VI) \geq g_a(d_a',\VI)$, the inequality \eqref{ineqn:ua_identified} holds. This completes the proof of the claim. The following theorem then follows immediately.

\begin{theorem} \label{theorem:optimal_estimation_strategy}
Suppose that the infection source adopts the infection strategy $\Lambda$ with safety margin $\sourced(\Lambda) = d_s$. If $\sourced = 0$, the estimation radius $\admind^*$ of the best-response estimation strategy is 0. If $ \sourced > 0$, the estimation radius of the best-response estimation strategy is given by
\begin{align}\label{admin_best_response}
\admind^* & = \begin{cases}
0, & \mbox{if } \admingain(d_s,\VI) < \admincost(\Vsp(\sourced)) -\admincost(\Vsp(0)), \\
\sourced, & \mbox{if } \admingain(d_s,\VI) > \admincost(\Vsp(\sourced)) -\admincost(\Vsp(0)), \\
0 \textrm{ or } \sourced, & \mbox{if } \admingain(d_s,\VI)  = \admincost(\Vsp(\sourced)) -\admincost(\Vsp(0)).
\end{cases}
\end{align}
\end{theorem}

We remind the reader that the quantities $\admingain(d_s,\VI)$ and $\Vsp(\sourced)$ in Theorem \ref{theorem:optimal_estimation_strategy} depend on the infection strategy $\Lambda$ only through the infection graph $\VI$ observed by the network administrator. Therefore, given the safety margin $\sourced$, the network administrator can formulate its best response using Theorem \ref{theorem:optimal_estimation_strategy} without knowing the source utility function.

We observe that if $d_s > 0$ in Theorem \ref{theorem:optimal_estimation_strategy}, then using an estimation radius of $d_a^*=0$ implies that the network administrator has zero probability of identifying the infection source. This happens when the reward of catching the infection source is significantly lower than the cost of probing more nodes. In practical systems, attempts should be made to keep the cost of probing each node in the network sufficiently small so that the infection source can be identified with positive probability. On the other hand, our result also points to the intuitive conclusion that for a source to escape identification with \emph{probability one}, the infection observation time $\tobs$ must be sufficiently long (cf.\ Theorem \ref{theorem:safety_margin_upper_bound}), and the source's safety margin must be chosen to be sufficiently large so that $\admincost(\Vsp(\sourced)) -\admincost(\Vsp(0))$ is large and the first case in \eqref{admin_best_response} holds.

\section{Best-response Infection Strategy for the Infection Source in a Tree}\label{sec:optimal_infection_strategy_trees}

In this section, we derive a best-response infection strategy for the infection source for the case where the underlying graph $G$ is a tree. We assume that the infection source knows the observation threshold $\nobs$ but not the observation time $\tobs$. We first derive our infection strategy based on the observation time $\tobs$, and show how to compute $\tobs$ from $\nobs$.

Given an estimation radius $\admind$ and an observation time $t$, the source designs a best-response infection strategy that maximizes its utility function. We first introduce the notion of a \emph{maximum infection strategy}. Let $\norm{\Lambda}$ denote the infection size of the infection strategy $\Lambda$.

\begin{definition} \label{def:maximum_infection_strategy} Given any safety margin $\sourced$ and observation time $t$, we define the maximum infection strategy with safety margin $\sourced$ to be the infection strategy that maximizes the number of infected nodes at time $t$ among all infection strategies that achieve the safety margin $\sourced$. Let the set of maximum infection strategies with safety margin $\sourced$ be $\MISSET{\sourced,t}$.
\end{definition}

For each estimation radius $d_a$, we design a best-response infection strategy $\Lambda^*$ for the infection source in three steps:
\begin{itemize}
  \item Step 1: Given any safety margin $\sourced$ and any observation time $t$, we find a maximum infection strategy $\MIS{\sourced,t} \in \MISSET{\sourced,t}$.
	\item Step 2: We search for the smallest $t$ such that $\norm{\MIS{\sourced,t}}\geq \nobs$, and set $\tobs=t$.			
  \item Step 3: Among all maximum infection strategies found in Step 2, we find one that maximizes the source's utility function as the best-response infection strategy, i.e., an infection strategy $\Lambda^* = \MIS{\sourced^*,\tobs}$ where
      \begin{align*}
      \sourced^* = \arg \max_{\sourced} u_s(d_a, \MIS{\sourced,\tobs}).
      \end{align*}
\end{itemize}
Note that under a given safety margin constraint $d_s$ and observation time $t$, the source's utility $u_s(\cdot,\cdot)$ is invariant to which infection strategy is chosen from $\MISSET{\sourced,t}$. In the following, we first determine the range of values that the safety margin $d_s$ can take for given set of maximum infection rates $\{\bar\lambda_m\}$, and observation time $t$, i.e., those values of $d_s$ such that $\MISSET{\sourced,t} \ne \emptyset$. We call such a safety margin \emph{feasible}. It is clear that $\MISSET{0,t} = \{\Lambda_{\max}\}$, where $\Lambda_{\max}$ is the infection strategy in which each node at distance $m$ from the source is infected at its respective maximum rate $\bar\lambda_{m-1}$. We next propose an algorithm to find an infection strategy in $\MISSET{\sourced,t}$, for all feasible $d_s > 0$.

\subsection{Maximum Infection Strategy with Safety Margin Constraint}

Given any observation time $t$, it turns out that not all values of $\sourced$ are feasible. To see this, consider an infection spreading along a linear network. Since the maximum number of hops the infection can spread from the source is $\dt$ (cf.\ \eqref{def:bard}), the safety margin cannot be more than $\lfloor \dt/2 \rfloor$. The following theorem provides the achievable upper bound for the safety margin in a tree. The proof is in Appendix \ref{appendix:theorem:safety_margin_upper_bound}.

\begin{theorem} \label{theorem:safety_margin_upper_bound}
Suppose the underlying graph $G$ is a tree. For any given observation time $t$, the largest feasible safety margin is
\begin{align}
\bar{\sourced} = \floor{\frac{\dt}{2}}. \label{eqn:initiation_criterion}
\end{align}
\end{theorem}

From Theorem \ref{theorem:safety_margin_upper_bound}, we see that the safety margin of any infection strategy can take values only from the set $[0, \bar{\sourced}]$. Given a feasible safety margin $\sourced$, we next show how to design a maximum infection strategy $\MIS{\sourced,t} \in \MISSET{\sourced,t}$. In the rest of this section, we adopt the following notations: Let $\GI[t]$ be the infection graph at observation time $t$ generated by a given infection strategy, which will be clear from the context. Let $T_u$ to be the subtree of $\GI[t]$ rooted at node $u$ with the first link in the path from $u$ to the source node $\sss$ removed. Let $\leaf{u}$ to be a leaf node in $T_u$ that has maximum distance from $u$, i.e.,
\begin{align}
		\leaf{u}\in \{i \in T_u : d(u,i)= \max_{j \in T_u} d(u,j)\}. \label{eqn:leaf}
\end{align}
We start by defining a dominant path and showing an elementary result related to this definition.
\begin{definition}
For any observation time $t$, a dominant path is a path between the source $v^*$ and any node in the infection graph $\GI[t]$ that has the maximum distance.
\end{definition}

\begin{lemma}\label{lemma:dominant_path}
Suppose that $G$ is a tree. Consider a maximum infection strategy with a feasible safety margin $\sourced$ that results in an infection graph $\GI[t]$ at time $t$. Then, the infection along each edge in any dominant path of $\GI[t]$ has maximum infection rate $\bar{\lambda}_m$ if the endpoint of the edge closer to the source is at distance $m$ from it.
\end{lemma}

The proof of Lemma \ref{lemma:dominant_path} is provided in Appendix \ref{appendix:lemma:dominant_path}. To get a safety margin $\sourced > 0$, the intuition is to construct one dominant path $\dpath$ starting at $v^*$ so that the Jordan center is biased towards the leaf node at the other end of $\dpath$, which in turn results in a safety margin $\sourced$. We discuss how to select an optimal dominant path in Algorithm \ref{algo:optimal_dominant_path}. For now, we assume that a dominant path $\dpath$ is given. Our proposed $\DIS[d_s,t]$ strategy, given in Strategy \ref{algo:DIS}, is defined by a set of parameters, $\{\lambda_m : m\in[0,\dt-1]\}$. Consider any infected node $i$ on the path $\dpath$. Suppose $d(\sss, i)=m$ and let $j$ be the susceptible neighboring node of $i$ on $\dpath$. The node $i$ infects $j$ with rate $\bar{\lambda}_m$ and infects all its other susceptible neighbors with rate $\lambda_m$. On the other hand, for any infected node that is not on $\dpath$, it infects all its susceptible neighbors with the same rate that it itself was previously infected.

In the following discussion, we show that if the parameter $\lambda_m$ is set to be that in \eqref{eqn:infection_rates}, and the dominant path is selected by Algorithm \ref{algo:optimal_dominant_path}, then $\DIS[d_s,t] \in \MISSET{d_s,t}$.

\floatname{algorithm}{Strategy}
\begin{algorithm}[!t]
\caption{Dominant Infection Strategy $\DIS[d_s,t]$}
\label{algo:DIS}
\begin{algorithmic}[1]
\STATE{\textbf{Inputs}: $G=(V,E)$, source $\sss$, observation time $t$,  required safety margin $\sourced$, set of maximum infection rates $\{\bar{\lambda}_m\}$, infection rates $\{\lambda_{m,j}\}$ given in \eqref{eqn:infection_rates}, and a dominant path $\dpath$ found by Algorithm \ref{algo:optimal_dominant_path} below.}
\STATE{\textbf{Output}: $\lambda(u,v)$, the infection rate for every $(u,v) \in \GI[t]$.}
\FOR{each $u \in V$}
	\IF{$u \in \dpath$}
		\STATE{Let $m = d(\sss, i)$.}
		\STATE{If $v$ is the susceptible neighbor of $u$ on $\dpath$, set $\lambda(u,v) = \bar{\lambda}_m$}.
		\STATE{For any other susceptible neighbor $v$ of $u$ not on $\dpath$, set $\lambda(u,v) = \lambda^{\DIS}_{m,j}$ where $j = d(u,v)$, and pass the message $(m,1)$ to $v$.}
	\ELSE
		\STATE{Let $(a_u,b_u)$ be the message received by $u$. For any susceptible neighbor $v$ of $u$, set $\lambda(u,v) = \lambda^{\DIS}_{a_u,b_u}$. Pass the message $(a_u,b_u+1)$ to $v$.}
	\ENDIF
\ENDFOR
\RETURN{$\{\lambda(u,v): (u,v)\in \GI[t]\}$}
\end{algorithmic}
\end{algorithm}

Suppose that $\sss$ has $k$ neighbors $v_1, v_2, \cdots, v_k$ in $\GI$, where $k \ge 2$. Without loss of generality, suppose that the labels are assigned so that $d(v,\leaf{v_1}) \geq d(v,\leaf{v_2}) \geq \ldots \geq d(v,\leaf{v_k})$. We have the following elementary result, the proof of which is provided in Appendix \ref{appendix:lemma:d1>d2}.

\begin{lemma}\label{lemma:d1>d2}
Suppose that $G$ is a tree. Given an infection graph $\GI[t]$, if $d(\sss,\leaf{v_1}) > d(\sss,\leaf{v_2})$, then at least one end of any diameter of $\GI[t]$ is in the subtree $T_{v_1}$.
\end{lemma}

\begin{figure}[!t]
  \centering
  \psfrag{a}[][][1][0]{$\ttree{u_{d(\sss,\leaf{v_1})-1}}$}
  \psfrag{e}[][][1][0]{$\ttree{u_m}$}
  \psfrag{c}[][][1][0]{$\ttree{u_0}$}
  \psfrag{4}[][][1][0]{$u_0$}
  \psfrag{5}[][][1][0]{$u_m$}
  \psfrag{6}[][][1][0]{$u_{\dt}$}
  \psfrag{1}[][][1][0]{$\sss$}
  \psfrag{3}[][][1][0]{$\leaf{v_1}$}
  \includegraphics[width=0.5\textwidth]{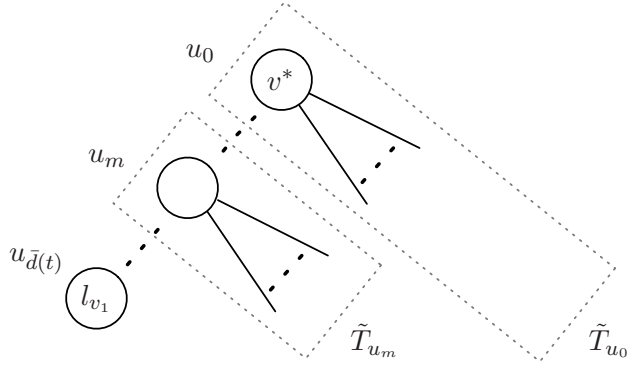}
  \caption{Illustration of the infection graph $\GI$.}
\label{fig:infection_network}
\end{figure}

We now show how to find the optimal parameter $\lambda_m$ in $\DIS[d_s,t]$ so that it achieves a safety margin $\sourced$. From Lemma \ref{lemma:dominant_path}, we have
\begin{align}
d(\sss, \leaf{v_1}) &= \dt. \label{eqn:dominant_path_length}
\end{align}
Let $\leaf{v_1}$ be a leaf node so that $\rho(\sss, \leaf{v_1}) = (u_0=v^*, u_1, \cdots, u_{\dt}=\leaf{v_1})$ is the given dominant path $\dpath$. For $0 \leq m \leq d(\sss,\leaf{v_1})-1$, define $\tilde{T}_{u_m}$ to be $T_{u_m}\backslash T_{u_{m+1}}$. Similar to the definition in \eqref{eqn:leaf}, let $\tleaf{u_m}$ be a leaf node in $\ttree{u_m}$ that has maximum distance from $u_m$. Figure \ref{fig:infection_network} shows an illustration of $\GI[t]$ and $\ttree{u_m}$.

For each $m\geq 1$, let
\begin{align}\label{tm}
t_m = \sum_{k=0}^{m-1}\bar{\lambda}_k^{-1}
\end{align}
be the time taken to infect node $u_m$. Suppose the nodes in $\rho(u_m,\tleaf{u_m})$ are infected at an \emph{average} rate $\lambda_m$ with
\begin{align}
d(u_m, \tleaf{u_m}) &= \left\lfloor \lambda_m (t-t_m) \right\rfloor. \label{eqn:u_m_path_length}
\end{align}

Let $\mathcal{D}(u_m)$ to be the set of longest paths such that one end of any path $D(u_m) \in \mathcal{D}(u_m)$ is $\leaf{v_1}$ and the other end is in $\tilde{T}_{u_m}$. From \eqref{eqn:dominant_path_length} and \eqref{eqn:u_m_path_length}, the number of vertices in $D(u_m)$ is
\begin{align}
|D(u_m)| &= d(\leaf{v_1},u_m) + d(u_m, \tleaf{u_m})+1 \nonumber \\
&= d(\sss, \leaf{v_1})-d(\sss, u_m) + d(u_m, \tleaf{u_m})+1 \nonumber \\
&= \dt - m + \left\lfloor \lambda_m(t-t_m)\right\rfloor +1. \label{eqn:u_m_diameter_length}
\end{align}
Let $\est(u_m)$ to be a node in the middle of $D(u_m)$, i.e., $\est(u_m) \in \arg \min_{v \in D(u_m)} \bar{d}(v, D(u_m))$. Since the infection is propagated at the maximum rates along $P_d$, we can always choose $\est(u_m) \in P_d$ with
\begin{align}
&d(\sss, \est(u_m)) \nonumber \\
&= d(\sss, \leaf{v_1}) - d(\est(u_m), \leaf{v_1}) \nonumber \\
&=d(\sss, \leaf{v_1}) - \left( \left\lceil \frac{|D(u_m)|}{2}\right\rceil -1 \right). \label{eqn:u_m_safety_margin_middle}
\end{align}
In order to maximize the number of infected nodes, we maximize $|D(u_m)|$. Note that $|D(u_m)|$ is odd, because otherwise we can always increase $\lambda_m$ so that we have one more node in $|D(u_m)|$, but \eqref{eqn:u_m_safety_margin_middle} remains the same. Then we have
\begin{align}
&d(\sss, \est(u_m)) \nonumber \\
&=d(\sss, \leaf{v_1}) - \left( \frac{|D(u_m)|}{2}+\frac{1}{2} -1 \right) \nonumber \\
&=\frac{1}{2}\left(\dt +m - \left\lfloor \lambda_m(t-t_m)\right\rfloor \right). \label{eqn:u_m_safety_margin}
\end{align}

Since $d_s > 0$, we must have $d(\sss,\leaf{v_1}) > d(\sss,\leaf{v_2})$. From Lemma \ref{lemma:d1>d2}, we see that one end of any diameter is a $\leaf{v_1}$. Then any diameter is in the set of paths $\bigcup_{m=0}^{\dt-1}\mathcal{D}(u_m)$ and the set of Jordan centers is a subset of $\{\est(u_m): D(u_m) \in \bigcup_{m=0}^{\dt-1}\mathcal{D}(u_m)\}$. The safety margin requirement $\sourced$ is satisfied if the right hand side of \eqref{eqn:u_m_safety_margin} has value at least $\sourced$ for every $m \in [0, d_s]$, i.e.,
\begin{align}
\left\lfloor \lambda_m(t-t_m)\right\rfloor \le h_m \triangleq \dt -2\sourced + m. \label{eqn:lambda_m_upper_bound}
\end{align}
To maximize the number of infected nodes at time $t$, we choose each $\lambda_m$ to be as large as possible. Therefore, for $m\in [0, d_s]$, we choose $\lambda_m$ to be the largest value so that equality holds in \eqref{eqn:lambda_m_upper_bound}, i.e.,
\begin{align}\label{eqn:average_infection_rates}
\lambda_m = \frac{h_m}{t-t_m}.
\end{align}
We then find $\delta_{t,m} \geq 0$ for $j \in [0,h_m]$ such that
\begin{align*}
\sum_{j\in A_m} \left(\bar\lambda_{m+j} - \delta_{t,m}\right)^{-1} = t - t_m - \sum_{j\notin A_m} \bar\lambda_{m+j}^{-1},
\end{align*}
where $A_m = \{j \leq h_m : \bar\lambda_{m+j} > \lambda_m\}$. Such a $\delta_{t,m}$ exists because $h_m \leq \dt - m$, the longest distance the infection can propagate from $u_m$. Finally, we let the infection at each node $v \in \ttree{u_m}$ such that $d(u_m, v) = j \leq h_m$ spread at rate $\bar\lambda_{m+j} - \delta_{t,m}$. If $m \in (d_s, \dt)$, we choose $\bar\lambda_{m+j}$ to be the spreading rate for all $v \in \ttree{u_m}$ such that $d(u_m, v) = j$. Note that with this choice, $\est(u_m)$ is the Jordan center on $\dpath$ for all $m\in [0, d_s]$.

In summary, for a $v \in \ttree{u_m}$ such that $d(u_m, v) = j$, we let it infect its susceptible neighbors not on the dominant path with rate
\begin{align}\label{eqn:infection_rates}
\lambda^{\DIS}_{m,j} &=
\begin{cases}
\bar\lambda_{m+j} - \delta_{t,m}, & \mbox{if } j \in A_m,\ \mbox{and } 0 \leq m \leq d_s, \\
\bar\lambda_{m+j}, & \mbox{otherwise}.
\end{cases}
\end{align}
We have the following result. The proof is provided in Appendix \ref{appendix:lemma:DIS_Pd}.

\begin{lemma}\label{lemma:DIS_Pd}
Suppose that $\DIS[d_s,t]$ with safety margin $d_s>0$ has the dominant path $\dpath$. Then, it maximizes the number of infected nodes at time $t$ amongst all infection strategies with safety margin $d_s$ and dominant path $\dpath$.
\end{lemma}

In the following, we show how to find the optimal dominant path. Given any dominant path $\dpath=(u_0=v^*,\ldots,u_{\dt})$ as an input of $\DIS[d_s,t]$ with safety margin $d_s>0$, we have from \eqref{eqn:infection_rates}, that the number of infected nodes is given by
\begin{align}\label{eqn:number_infected_Pd}
\dt + \sum_{m=0}^{\dt-1} |\tilde{T}_{u_m}|.
\end{align}
To find an optimal dominant path so that the above sum is maximized, we use the procedure in Algorithm~\ref{algo:optimal_dominant_path}.

\floatname{algorithm}{Algorithm}
\setcounter{algorithm}{0}
\begin{algorithm}[!t]
\caption{Bellman-Ford Dominant Path Finding}
\label{algo:optimal_dominant_path}
\begin{algorithmic}[1]
	\STATE Perform a breadth-first search starting at $v^*$, and for each edge $(u,w)$ where $d(v^*,w) = d(v^*,u)+1 \leq \dt$, assign the following weight:
	\begin{align*}
	w(u,w) = \left| \{ v \in T_u\backslash T_w :  d(v,u) \leq h(u) \} \right|,
	\end{align*}
	with
	\begin{align*}
	h(u) & =
	\begin{cases}
	\dt - 2d_s + d(v^*,u), & \mbox{if } d(v^*,u) \leq d_s, \\
	\dt - d(v^*,u), & \mbox{otherwise}.
	\end{cases}
	\end{align*}
	
	\STATE Use the Bellman-Ford algorithm \cite{Cormen2001} to find a maximal weighted path $(u_0,\ldots,u_{\bar{d}(t)})$ starting at $u_0=v^*$. The maximal weighted path is output as the dominant path, and its weight added to $\dt$ is output as $\norm{\DIS[d_s,t]}$.
\end{algorithmic}
\end{algorithm}

In Algorithm~\ref{algo:optimal_dominant_path}, since the weight we have assigned to each edge $(u_m, u_{m+1})$ in the maximal weight path found corresponds exactly to $|\tilde{T}_{u_m}|$ in \eqref{eqn:number_infected_Pd}, the algorithm gives us the optimal dominant path. In the first step of Algorithm \ref{algo:optimal_dominant_path}, the weights $w(u,w)$ for all neighbors $w$ of $u$ in $T_u$ can be found by performing another breadth-first search in the tree $T_u$. The time complexity of the first step is thus $O(n^2)$,\footnote{A function is said to be $O(f(n))$ if it is upper bounded by $kf(n)$ for some constant $k>0$ and for all $n$ sufficiently large.}  where $n$ is the number of vertices within a distance $\dt$ of $v^*$ \cite{Cormen2001}. The Bellman-Ford algorithm in the second step also has time complexity $O(n^2)$. Therefore, the overall time complexity of Algorithm \ref{algo:optimal_dominant_path} is $O(n^2)$.

Lemma \ref{lemma:DIS_Pd} and Algorithm \ref{algo:optimal_dominant_path} then lead to the following result.

\begin{theorem}\label{theorem:DIS}
Suppose $G$ is a tree. For any observation time $t$ and feasible safety margin $d_s$, $\DIS[d_s,t] \in \MISSET{d_s,t}$ if the dominant path is found by Algorithm~\ref{algo:optimal_dominant_path}.
\end{theorem}

Since $\norm{\DIS[d_s,t]}$ is non-decreasing in $t$, we can now perform a binary search procedure in Algorithm~\ref{algo:binarysearch} to determine the smallest $t$ such that $\norm{\DIS[d_s,t]} \geq \nobs$. Note that it suffices to perform the binary search over the times $\{t_m : m\geq 1\}$ defined in \eqref{tm}, because for any $t \in [t_m, t_{m+1})$, $m\geq 1$, we have $\norm{\DIS[d_s,t]} = \norm{\DIS[d_s,t_m]}$ as the right hand side of \eqref{eqn:lambda_m_upper_bound} remains unchanged. Let
\begin{align}
&x_0 = \min \{m : |\{v:d(v^*,v) \leq \bar{d}(t_m)\}| \geq \nobs\}, \label{x0}\\
&y_0 = \min \{m : |\{v:d(v^*,v) \leq \bar{d}(t_m)-2d_s\}| \geq \nobs\}. \label{y0}
\end{align}
We initialize the search to be over $[t_{x_0}, t_{y_0}]$. Because not all vertices within distance $\bar{d}(\tobs)$ are infected by $\DIS[d_s,\tobs]$, while all nodes within distance $\bar{d}(\tobs)-2d_s$ are infected, we have $\norm{\DIS[d_s,t_{x_0}]} \leq \nobs \leq \norm{\DIS[d_s,t_{y_0}]}$. The binary search takes at most $O(\log \nobs)$ search steps. Assuming that the tree $G$ has bounded degree $\beta$, in each search step the computation of $\norm{\DIS[d_s,t_m]}$ using Algorithm~\ref{algo:optimal_dominant_path} takes at most $O(\beta^{4d_s}\nobs^2)$ time complexity. Therefore the overall time complexity to find $\DIS[d_s,\tobs]$ is $O(\beta^{4d_s}\nobs^2 \log\nobs)$.

\begin{algorithm}[!htb]
\caption{Binary Search for $\tobs$}
\label{algo:binarysearch}
\begin{algorithmic}[1]
	\STATE Initialize $x = x_0$ using \eqref{x0}, and $y = y_0$ using \eqref{y0}.
	\WHILE{$x < y-1$}
	\STATE Set $m = \ceil{(x + y)/2}$.
	\IF{$\norm{\DIS[d_s,t_m]} \leq \nobs$}
	\STATE Set $x = m$.
	\ELSE
	\STATE Set $y = m$.
	\ENDIF
	\ENDWHILE
	\RETURN Output $\tobs = t_y$.
\end{algorithmic}
\end{algorithm}

\subsection{Homogeneous Infection Rate Bounds}

If $\bar\lambda_m = \bar\lambda$ for all $m\geq 0$, then $\dt =\lfloor\bar{\lambda}t \rfloor$, and it can be shown from \eqref{eqn:infection_rates} that for all $m$ and $j$,
\begin{align}
\lambda^{\DIS}_{m,j} = \bar{\lambda} \cdot \min\left\{1, \frac{\lfloor\bar{\lambda}t \rfloor -2\sourced+m}{\bar{\lambda}t-m}\right\}, \label{eqn:homogenous_infection_rates}
\end{align}
i.e., the same rate is used to infect the vertices in the subtree $\tilde{T}_{u_m}$. In this case, the $\DIS[d_s,t]$ strategy need not pass additional distance information along with the infection.

\subsection{Infinite Regular Trees}

\begin{figure}[!t]
  \centering
  \psfrag{a}[][][0.8][0]{$\sss$}
  \psfrag{b}[][][0.8][0]{$\JC$}
  \psfrag{c}[l][][0.8][0]{$t-\sourced$}
  \psfrag{d}[l][][0.8][0]{$\lfloor t/2 \rfloor$}
  \psfrag{e}[][][0.8][0]{$\sourced$}
  \includegraphics[width=0.4\textwidth]{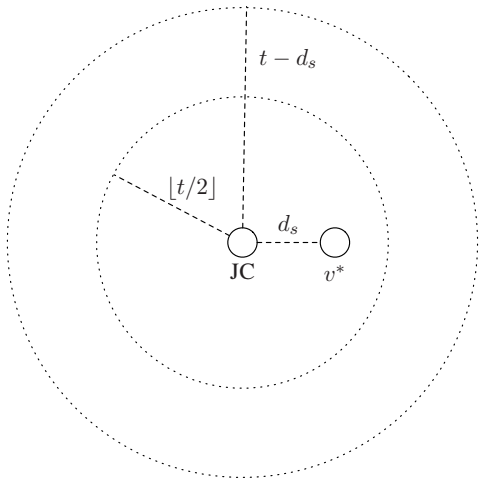}
  \caption{Illustration of the set of infected nodes by DIS and AD, where $\JC$ denotes the Jordan center of the set of infected nodes at time $t$. The set of infected nodes by DIS are all the nodes within the circle with radius $t-\sourced$, while the set of infected nodes by AD are all the nodes within the circle with radius $\lfloor t/2 \rfloor$. Since $\sourced$ is upper bounded by $\lfloor t/2 \rfloor$, we have $t-\sourced \ge \lfloor t/2 \rfloor$.}
\label{fig:infected_nodes_illustraction_DIS_AD}
\end{figure}

In the following, we consider the special case where the underlying network is an infinite $r$-regular tree, every node has $r>2$ neighboring nodes, and $\bar{\lambda}_m=1$ for all $m \ge 0$. The fastest infection strategy $\Lambda_{\max}$ is the one that sets all infection rates to be the upper bound 1. Then the number of nodes infected by the fastest infection strategy by time $t$ can be shown to be given by
\begin{align*}
|\VI(\Lambda_{\max})| = \frac{r(r-1)^t-2}{r-2}.
\end{align*}
This infection strategy has safety margin 0. Now suppose that the source wishes to achieve a safety margin $\sourced \leq \bar\sourced$ in \eqref{eqn:initiation_criterion}, the set of infected nodes by time $t$ by our proposed $\DIS[d_s,t]$ strategy can be shown to be all nodes with distance not greater than $t-\sourced$ from the Jordan center as shown in Fig. \ref{fig:infected_nodes_illustraction_DIS_AD}. Then, the number of nodes infected by time $t$ is
\begin{align*}
|\VI(\DIS[d_s,t])| = \frac{r(r-1)^{t-\sourced}-2}{r-2}.
\end{align*}
When $\sourced$ increases, the radius $t-\sourced$ decreases, and the number of nodes infected by the maximum infection strategy decreases. This is the necessary trade-off between the faster infection spreading speed and the larger safety margin of the source.

The paper \cite{Fanti2014} proposes a messaging protocol called adaptive diffusion (AD) under the assumption that the network administrator utilizes a ML estimation strategy. AD is a stochastic infection strategy, where its safety margin falls in the range $[1,\lfloor t/2 \rfloor]$ with probability one. Given any safety margin $\sourced \in [1,\lfloor t/2 \rfloor]$, with probability one, the set of infected nodes by time $t$ by AD can be shown to be all nodes with distance not greater than $\lfloor t/2 \rfloor$ from the Jordan center as shown in Fig. \ref{fig:infected_nodes_illustraction_DIS_AD}. Then, the number of nodes $|\VI(\text{AD})|$ infected by AD by time $t$ satisfies the following bound with probability one:
\begin{align*}
|\VI(\text{AD})| \le \frac{r(r-1)^{t-\lfloor t/2 \rfloor}-2}{r-2} \le |\VI(\DIS[d_s,t])|.
\end{align*}
Therefore, our proposed DIS strategy infects at least as many nodes as the AD strategy almost surely.

\subsection{Best-response Infection Strategy for the Infection Source}
Theorem \ref{theorem:DIS} shows how to find a maximum infection strategy for a feasible $\sourced$. We next identify one that maximizes the utility function of the infection source. We first present the following relationship between maximum infection strategies of different safety margins.

\begin{lemma} \label{lemma:maximum_infection_strategy_relationship}
Suppose $G$ is a tree. For any two safety margins $\sourced$ and $\sourced'$, where $0 \le \sourced < \sourced'  \le \bar{\sourced}$, any strategy $\MIS{\sourced} \in \MISSET{\sourced,\tobs}$ infects more nodes than any strategy $\MIS{\sourced'} \in \MISSET{\sourced',\tobs}$, i.e., $|\VI(\MIS{\sourced})| > |\VI(\MIS{\sourced'})|$.
\end{lemma}

\begin{IEEEproof}
From \eqref{eqn:average_infection_rates}, the average infection rate $\lambda_m$ is a non-increasing function of the safety margin. Since $\sourced < \sourced'$, the infection rate $\lambda_m$ in $\MIS{\sourced}$ is larger than or equal to that in $\MIS{\sourced'}$ for any $0 \le m \le \dt-1$. However, equality does not hold for all $m$, because otherwise, $\MIS{\sourced}$ and $\MIS{\sourced'}$ lead to the same infection graph, which in turn implies that $\sourced = \sourced'$, a contradiction. As a result, we have $|\VI(\MIS{\sourced})| > |\VI(\MIS{\sourced'})|$, which completes the proof of Lemma \ref{lemma:maximum_infection_strategy_relationship}.
\end{IEEEproof}

We now derive the best-response infection strategy based on Lemma \ref{lemma:maximum_infection_strategy_relationship}. Let $\MIS{d_s} \in \MISSET{d_s,\tobs}$ be any maximum infection strategy.

Assume that the network administrator uses the estimation radius $\admind$. We first consider the case where $\admind \ge \bar{\sourced}$. Since no infection strategies have safety margins greater than $\bar{\sourced}$, the inequality $\admind \ge \sourced$ always holds. As a result, the cost $\sourcecost(d_a)$ is always incurred. Therefore, to maximize its utility function, the infection source maximizes its reward $\sourcegain |\VI|$ by maximizing the number of infected nodes $|\VI|$. Lemma \ref{lemma:maximum_infection_strategy_relationship} then leads to the conclusion that $\MIS{0}=\Lambda_{\max}$ is a best-response infection strategy.

Next, consider the case where $\admind < \bar{\sourced}$. We claim that a best-response infection strategy is either $\MIS{0}$ or $\MIS{\admind+1}$. Following Theorem \ref{theorem:DIS}, it suffices to prove the claim by showing the following inequalities:
\begin{align}
u_s(\admind, \MIS{0}) > u_s(\admind, \MIS{\sourced'}), \ \ \ &\forall \ \sourced' \in (0, \admind]; \label{ineqn:us_identified}\\
u_s(\admind, \MIS{\admind+1}) > u_s(\admind, \MIS{\sourced'}), \ \ \ &\forall \ \sourced' \in (\admind+1, \bar{\sourced}]. \label{ineqn:us_not_identified}
\end{align}

We first show the inequality \eqref{ineqn:us_identified}. When $0 <\sourced' \le \admind$, the source incurs a cost of $\sourcecost(\admind)$ for both $\MIS{0}$ and $\MIS{\sourced'}$. In addition, from Lemma \ref{lemma:maximum_infection_strategy_relationship}, we have $|\VI(\MIS{0})| > |\VI(\MIS{\sourced'})|$, which in turn shows that the inequality \eqref{ineqn:us_identified} holds. We next show the inequality \eqref{ineqn:us_not_identified}. When $\admind+1 <\sourced' \le \bar{\sourced}$, the source does not incur a cost for both $\MIS{\admind+1}$ and $\MIS{\sourced'}$. From Lemma \ref{lemma:maximum_infection_strategy_relationship}, we have $|\VI(\MIS{\admind+1})|>|\VI(\MIS{\sourced'})|$, which shows that the inequality \eqref{ineqn:us_not_identified} holds. This completes the proof for the claim. The following theorem now follows immediately.

\begin{theorem} \label{theorem:optimal_infection_strategy}
Suppose that $G$ is a tree. Then, for any estimation radius $\admind \geq 0$, a best-response infection strategy for the infection source is given by
\begin{align*}
\Lambda^* & = \begin{cases}
\MIS{0}, & \mbox{if } \sourcecost(d_a) < \sourcegain (|\VI(\MIS{0})|-|\VI(\MIS{\admind+1})|), \\
\MIS{\admind+1}, & \mbox{if } \sourcecost(d_a) > \sourcegain (|\VI(\MIS{0})|-|\VI(\MIS{\admind+1})|), \\
\MIS{0} \textrm{ or } \MIS{\admind+1}, & \mbox{if } \sourcecost(d_a) = \sourcegain (|\VI(\MIS{0})|-|\VI(\MIS{\admind+1})|), \\
\end{cases}
\end{align*}
where $\MIS{d} \in \MISSET{d,\tobs}$ for all $d \geq 0$.
\end{theorem}

\section{Nash Equilibrium in a Tree}\label{sec:nash_equilibrium}

In this section, we derive conditions under which a Nash equilibrium for the strategic game played by the network administrator and the infection source exists. We also derive explicitly their respective strategies at these Nash equilibria.

\begin{theorem} \label{theorem:nash_equilibrium}
Suppose $G$ is a tree, and $\bar{\sourced}$ in \eqref{eqn:initiation_criterion} is greater than 0. Then, the strategic game of infection spreading and source identification \eqref{eqn:admin_utility}-\eqref{eqn:source_utility} has the following properties:
\begin{enumerate}[(a)]
  \item \label{theorem:nash_equilibrium:00} Let $\MIS{0}\in\MISSET{0,\tobs}$. The strategy pair $(0,\MIS{0})$ is a Nash equilibrium iff $u_s(0, \MIS{1}) \le u_s(0, \MIS{0})$, i.e., $\sourcecost(0) \le \sourcegain \left( |\VI(\MIS{0})| - |\VI(\MIS{1})| \right)$ for any $\MIS{1}\in \MISSET{1, \tobs}$.
  \item \label{theorem:nash_equilibrium:01} For each $\MIS{1}\in\MISSET{1,\tobs}$, the strategy pair $(0,\MIS{1})$ is a Nash equilibrium iff $u_s(0, \MIS{0}) \le u_s(0, \MIS{1})$ and $u_a(1, \MIS{1}) \le u_a(0, \MIS{1})$, i.e., $\sourcecost(0) \ge \sourcegain \left( |\VI(\MIS{0})| - |\VI(\MIS{1})| \right)$ and $\admingain(1,\VI(\MIS{1})) \le \admincost(\Vsp(1)) - \admincost(\Vsp(0))$.
  \item \label{theorem:nash_equilibrium:no}No other pure strategy Nash equilibria exist.
\end{enumerate}
Furthermore, if $\admingain(\admind,\VI(\Lambda)) = \admingain(\admind)$ is non-increasing in $d_a$ for all infection strategies $\Lambda$, and $\admincost(\Vsp(\admind)) = \admincost(\admind)$ is non-decreasing in $d_a$ for all infection strategies $\Lambda$, then the sum utility of the two players is maximized at the strategy pairs $(0,\MIS{0})$ or $(0,\MIS{1})$ for all $\MIS{1} \in \MISSET{1,\tobs}$.
\end{theorem}

The proof of Theorem \ref{theorem:nash_equilibrium} is provided in Appendix \ref{appendix:theorem:nash_equilibrium}. From Theorem \ref{theorem:nash_equilibrium}, when a Nash equilibrium exists, whether a infection source is identified with positive probability or not depends on the relative gains and costs of the two players. It is interesting to note that if a Nash equilibrium exists, then the strategy of the network administrator in equilibrium has $d_a=0$, which corresponds to the Jordan center estimator. This shows that under the technical conditions given in Theorem \ref{theorem:nash_equilibrium}, the natural infection source estimator to use is the Jordan center estimator, instead of probing a neighborhood set of the Jordan centers.

\section{Simulation Results} \label{sec:simulation_results}

In this section, we present simulation results to evaluate the performance of our proposed infection and estimation strategies. We first compare DIS with AD, and then show the behavior of the best-response infection and estimation strategies under different gains and costs. For simplicity, our simulations are performed assuming that $g_a(\cdot,\cdot)=g_a$, $\admincost(\Vsp(\admind))=\admincost |\Vsp(\admind)|$, and $c_s(d_a)=c_s$.

\subsection{Extension to General Networks}\label{subsec:extension_general}
Although the paths along which the infection spreads from the source node forms a tree that is a subgraph of the given graph $G$, finding the best underlying tree over which to perform the infection spreading is a NP-hard problem (similar to the procedure used in Algorithm~\ref{algo:optimal_dominant_path}, this is equivalent to the longest path problem in a weighted graph, which is known to be NP-hard \cite{Cormen2001}).
To adapt our proposed DIS strategy for general networks, we adopt a heuristic: we first find a breadth-first search tree rooted at the infection source and then apply the DIS strategy on this tree. In the following, we show simulation results to verify the performance of the DIS strategy in general networks.

\subsection{Number of Infected Nodes}\label{subsec:number_of_infected}
We first evaluate the effectiveness of our proposed DIS algorithm in infecting nodes. We perform simulations on four kinds of networks: random trees where each node has a degree uniformly drawn from the set $\{2, 3\}$, scale-free networks \cite{Barabasi2009} with 5000 nodes, the western states power grid network of the United States \cite{Watts1998} containing 4941 nodes, and a part of the Facebook network with 4039 nodes \cite{McAuley2012}.

The benchmark we compare against is the AD infection strategy proposed in \cite{Fanti2014}, where AD is shown to be order-optimal for the source for infinite regular trees (with heuristic extensions to more general networks). We let $t$ to be even and $\bar{\lambda}_m=1$ for $m \ge 0$ in the simulations in order not to conflate the effect of the rate bounds with the other factors. Given any observation time $\tobs$, the safety margin $d_s(\text{AD})$ resulting from the AD falls in the range $[1,\tobs/2]$ with probability one, and the set of infected nodes at time $\tobs$ is
\begin{align*}
\VI(\text{AD}) = \{ u \in G: d(\tilde{v},u) \le \tobs/2 \},
\end{align*}
where $\tilde{v}$ is picked uniformly at random from the set of nodes in $G$ with distance $d_s(\text{AD})$ from the infection source $\sss$.

\begin{figure}[!t]
  \centering
  \psfrag{a}[l][][0.65][0]{Random trees, DIS}
  \psfrag{b}[l][][0.65][0]{Random trees, AD}
  \psfrag{c}[l][][0.65][0]{Power grid, DIS}
  \psfrag{d}[l][][0.65][0]{Power grid, AD}
  \psfrag{n}[l][][0.65][0]{Scale-free, DIS}
  \psfrag{u}[l][][0.65][0]{Scale-free, AD}
  \psfrag{e}[l][][0.65][0]{Facebook, DIS}
  \psfrag{k}[l][][0.65][0]{Facebook, AD}
  \psfrag{g}[][][0.8][0]{Average infection size}
  \psfrag{h}[][][0.8][0]{Safety margin $d_s$}
  \includegraphics[width=0.5\textwidth]{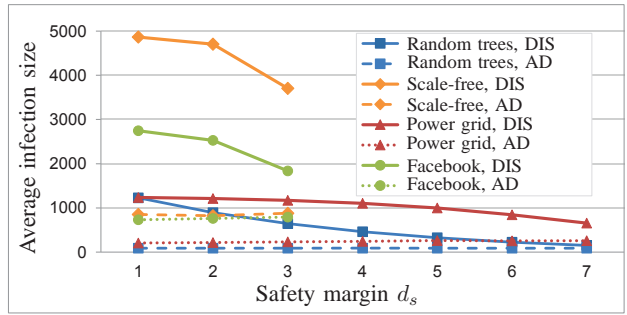}
  \caption{Average numbers of infected nodes for DIS and AD for different networks.}
\label{fig:average_infection_size_DIS_AD}
\end{figure}

We let the observation time $\tobs$ to be $14$, $14$, $6$ and $6$ for random trees, the power grid network, scale-free networks and the Facebook network, respectively. The observation times for scale-free networks and the Facebook network are chosen to be relatively small because these networks are highly connected and the average distances between each pair of nodes in the scale-free network and the Facebook network are only 4.6 and 5.5 hops, respectively. The safety margin requirement $d_s$ is set to be $1,2,\ldots,\tobs/2$, respectively. We run 1000 simulation runs for each kind of network and each value of $d_s$. Fig. \ref{fig:average_infection_size_DIS_AD} shows the average number of infected nodes for both DIS and AD. As expected, we see that there is a trade off for DIS between the number of infected nodes and the safety margin. We also see that DIS consistently infects more nodes than AD.

\subsection{Best-response Infection Strategy}
\begin{figure}[!t]
 \centering
  \psfrag{a}[l][][0.65][0]{low $\sourcecost$}
  \psfrag{b}[l][][0.65][0]{medium $\sourcecost$}
  \psfrag{c}[l][][0.65][0]{high $\sourcecost$}
  \psfrag{d}[][][0.8][0]{Average utility}
  \psfrag{e}[][][0.8][0]{Estimation radius $\admind$}
  \subfigure[Random trees.]{
    \label{fig:infection_strategy_tree}
    \includegraphics[width=0.45\textwidth]{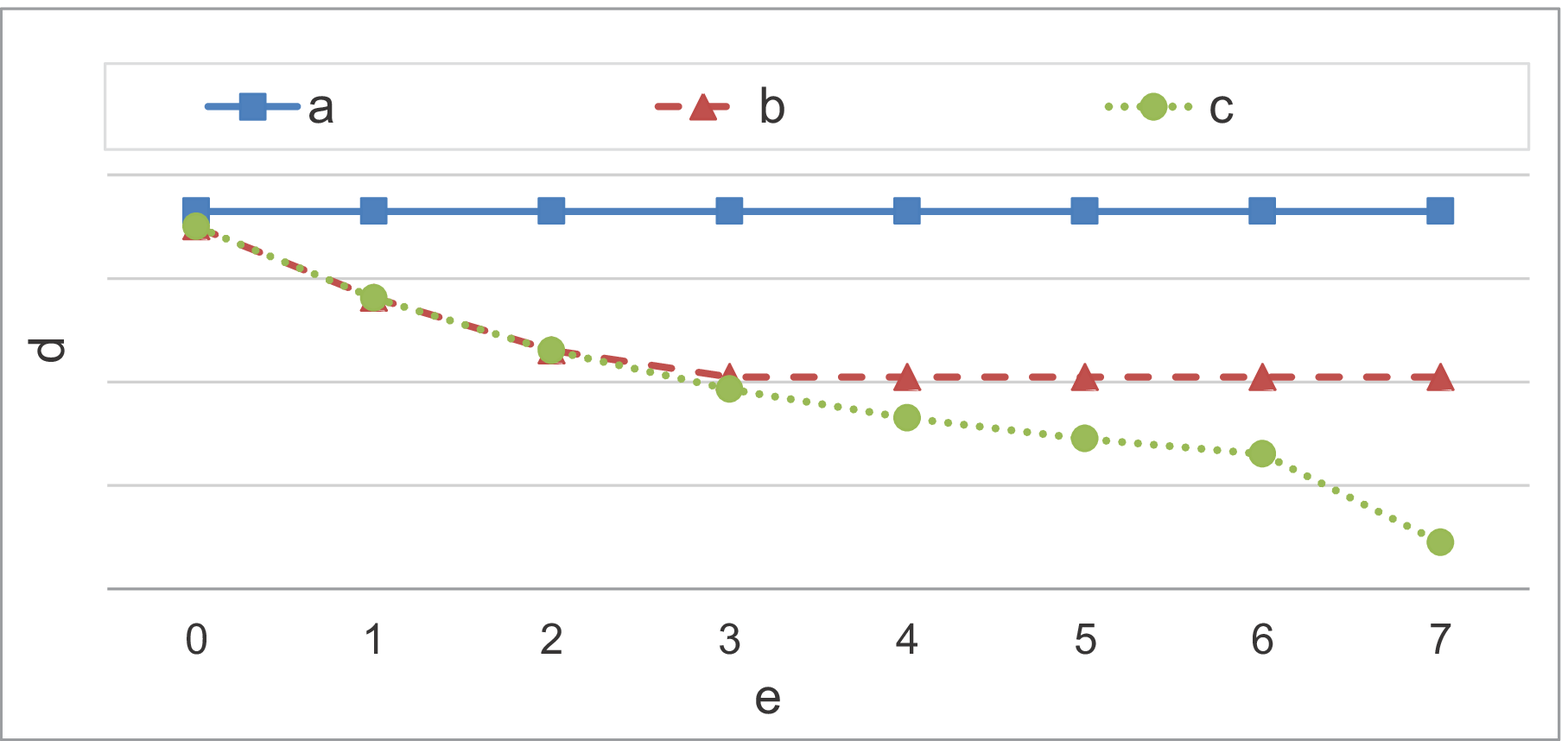}}
  \hspace{0.1in}
  \subfigure[Facebook network.]{
    \label{fig:infection_strategy_facebook}
    \includegraphics[width=0.45\textwidth]{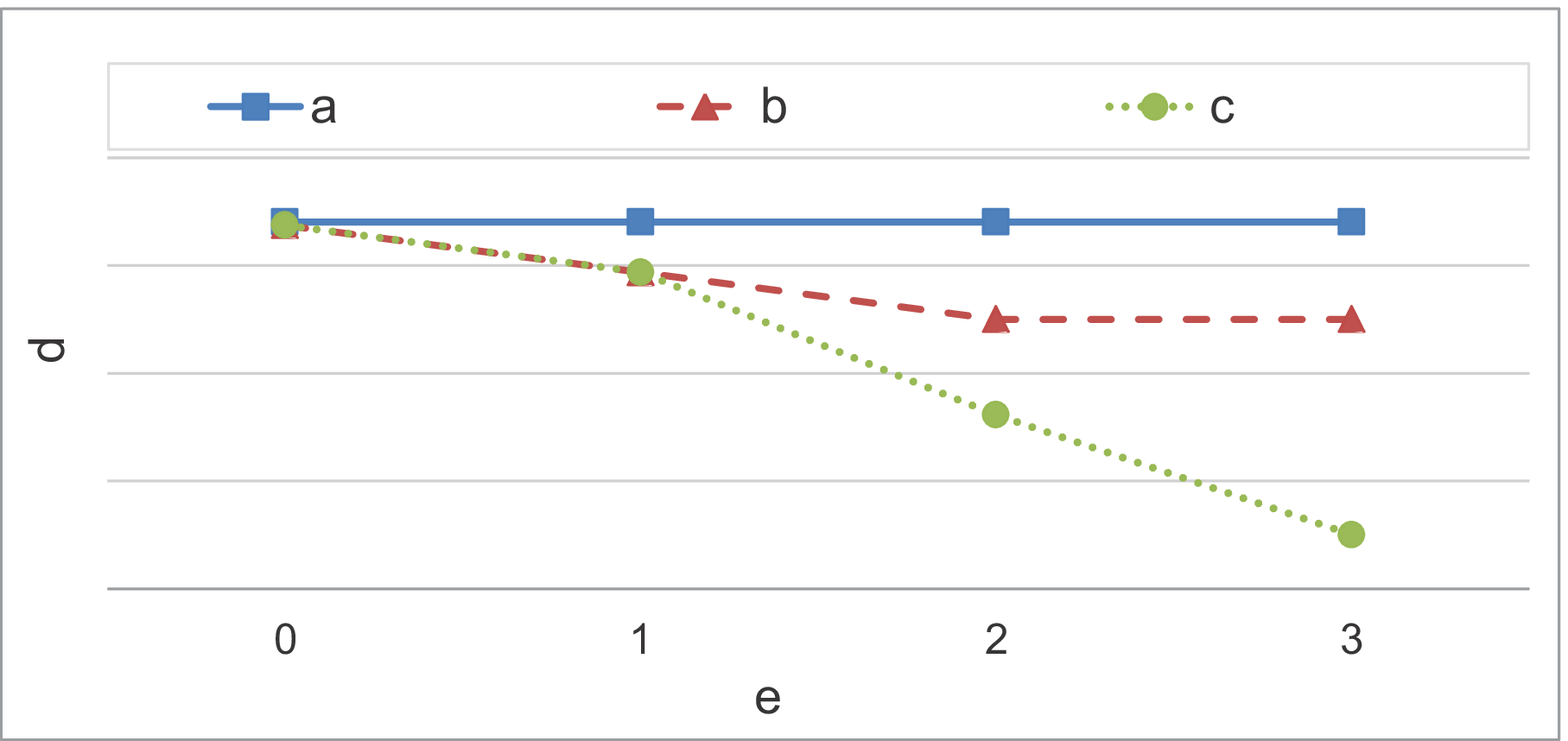}}
\caption{Average utility of the best-response infection strategies for the infection source. When $\sourcecost$ is low, $\DIS[0,\tobs]$ is the best-response infection strategy for all $\admind$. When $\sourcecost$ is high, $\DIS[\admind+1,\tobs]$ are the best-response infection strategies for $\admind < \bar{\sourced}$, and $\DIS[0,\tobs]$ is the best-response infection strategy for $\admind = \bar{\sourced}$. When $\sourcecost$ is medium, $\DIS[\admind+1,\tobs]$ are the best-response infection strategies for $\admind \le 2$ for random trees and $\admind \le 1$ for Facebook network, respectively, and $\DIS[0,\tobs]$ is the best-response infection strategy for other values of $\admind$.}
\label{fig:infection_strategy}
\end{figure}

We then evaluate the proposed best-response infection strategy on random trees and the Facebook network. The observation time $\tobs$ for each network is chosen as in Section \ref{subsec:number_of_infected}. For each value of $\admind \in [0, \bar{\sourced}]$, where $\bar{\sourced} = \tobs/2$, the best-response infection strategy is given by Theorem \ref{theorem:optimal_infection_strategy}. We fix the gain $\sourcegain$ for all cases and vary the cost $\sourcecost$ to make it low, medium and high, compared to $\sourcegain$. Specifically, we set $\sourcecost$ to be 400$\sourcegain$, 1200$\sourcegain$ and 2000$\sourcegain$ for random trees, and 50$\sourcegain$, 500$\sourcegain$ and 1500$\sourcegain$ for Facebook network. Let $\DIS_{\sourced}$ denote the DIS strategy with safety margin constraint $\sourced$. We run 1000 simulation runs for each setting and plot the average utility of the best-response infection strategies for the infection source in Fig. \ref{fig:infection_strategy}. We observe similar trends from Fig. \ref{fig:infection_strategy} for both random trees and the Facebook network, even though the DIS strategy was derived for tree networks.
 \begin{itemize}
   \item When the cost $\sourcecost$ is low compared to the gain $\sourcegain$, the infection source always chooses $\DIS[0,\tobs]$ as its infection strategy. As a result, the infection source is identified by the network administrator. However, it maximizes its reward by infecting the most number of nodes.
   \item When the cost $\sourcecost$ is high compared to the gain $\sourcegain$, the infection source chooses $\DIS[\admind+1,\tobs]$ for $\admind < \bar{\sourced}$ to ensure that the network administrator does not identify it, and chooses $\DIS[0,\tobs]$ for $\admind = \bar{\sourced}$ as it can not find any infection strategy with a safety margin greater than $\bar{\sourced}$ (cf.\ Theorem \ref{theorem:safety_margin_upper_bound}).
   \item When the cost $\sourcecost$ is medium compared to the gain $\sourcegain$, the infection source chooses $\DIS[\admind+1,\tobs]$ when $\admind$ is small. As $\admind$ increases, the infection source switches to $\DIS[0,\tobs]$ as its infection strategy.
 \end{itemize}

\subsection{Best-response Estimation Strategy}\label{subsec:bes_estimation}
\begin{figure}[!t]
 \centering
  \psfrag{a}[l][][0.65][0]{low $\admingain$}
  \psfrag{b}[l][][0.65][0]{medium $\admingain$}
  \psfrag{c}[l][][0.65][0]{high $\admingain$}
  \psfrag{d}[][][0.8][0]{Average utility}
  \psfrag{e}[][][0.8][0]{Safety margin $\sourced$}
  \subfigure[Random trees.]{
    \label{fig:estimation_strategy_tree}
    \includegraphics[width=0.45\textwidth]{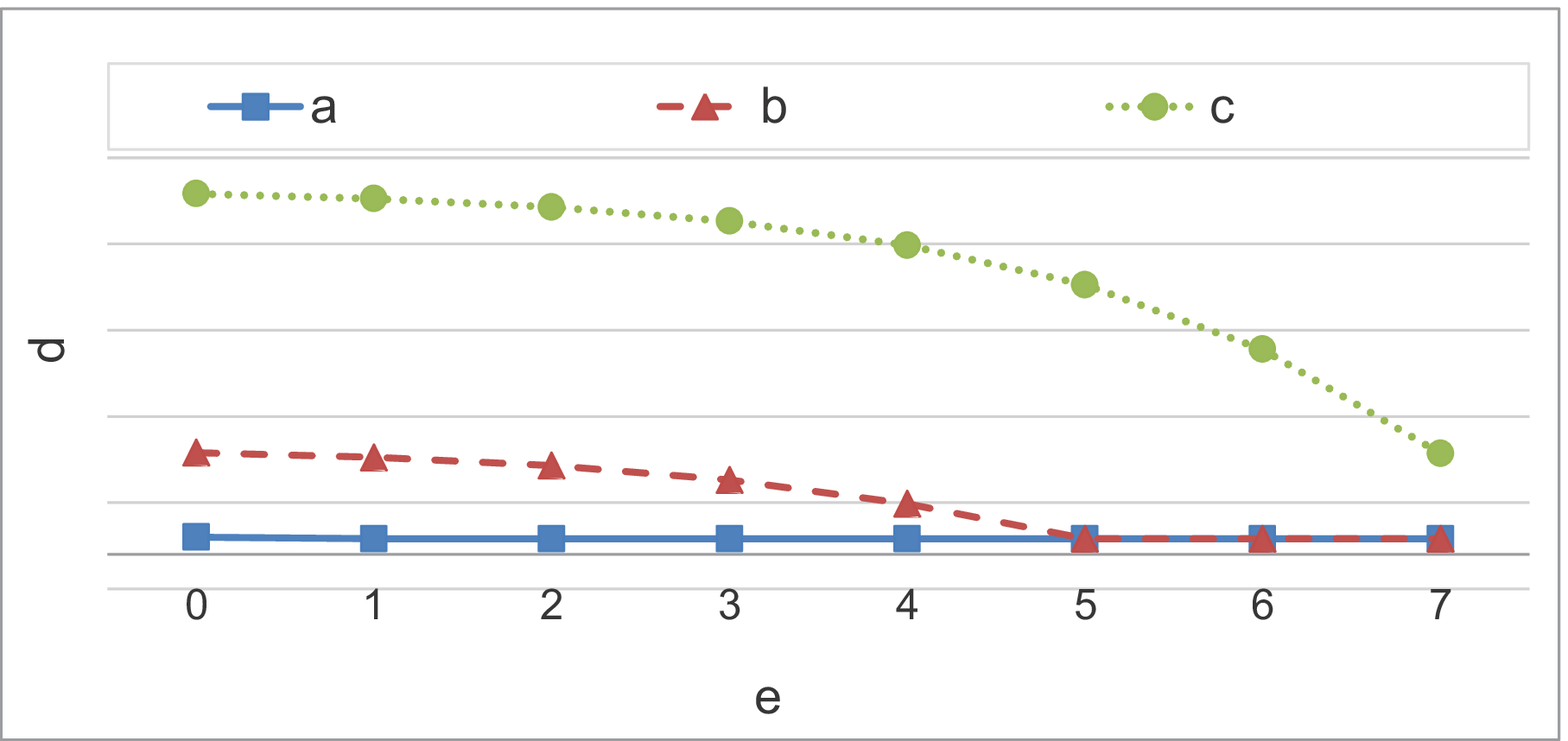}}
  \hspace{0.1in}
  \subfigure[Facebook network.]{
    \label{fig:estimation_strategy_facebook}
    \includegraphics[width=0.45\textwidth]{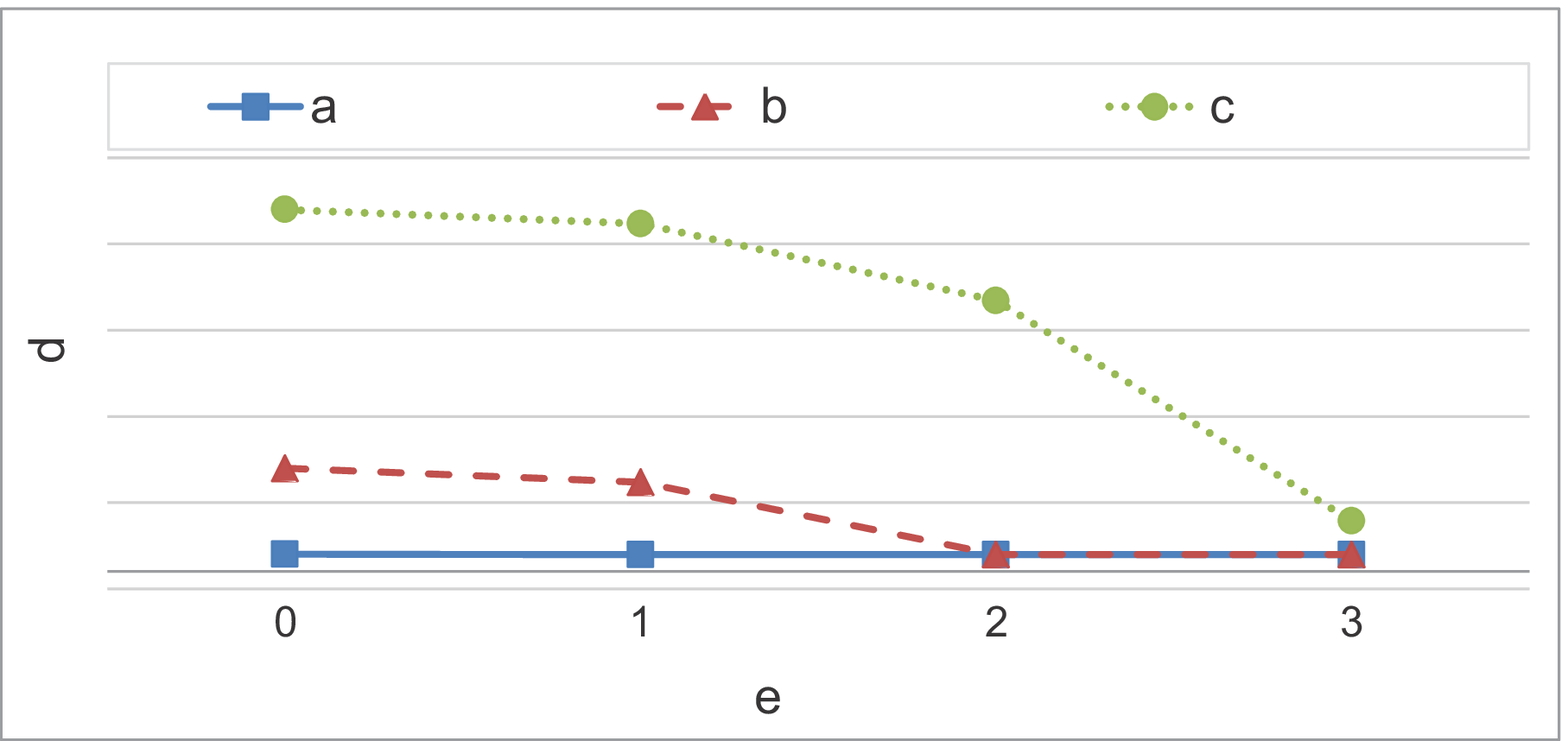}}
\caption{Average utility of the best-response estimation strategies for the network administrator. When $\admingain$ is low, $\admind$ is chosen to be $0$ for all $\sourced$. When $\admingain$ is high, $\admind$ is chosen to be $\sourced$ for all $\sourced$. When $\admingain$ is medium, $\admind$ is set to be $\sourced$ for $\sourced \le 4$ for random trees and $\sourced \le 1$ for Facebook network, respectively, and $\admind$ is set to be $0$ for other values of $\sourced$.}
\label{fig:estimation_strategy}
\end{figure}

Lastly, we evaluate the proposed best-response estimation strategy on random trees and Facebook network. Given any infection strategy $\Lambda$ with safety margin $\sourced(\Lambda) = d_s$, where $\sourced \in [1,\bar{\sourced}]$ and $\bar{\sourced} = \tobs/2$, the best-response estimation strategy is given by Theorem \ref{theorem:optimal_estimation_strategy}. We choose $\nobs$ so that it corresponds to the same observation time $\tobs$ used for each network in Section \ref{subsec:number_of_infected}. We fix the cost $\admincost$ for all cases and vary the gain $\admingain$ to make it low, medium and high, compared to $\admincost$. Specifically, we set $\admingain$ to be $\admincost$, 50$\admincost$ and 200$\admincost$ for random trees, and $\admincost$, 500$\admincost$ and 2000$\admincost$ for Facebook network. We run 1000 simulation runs for each setting and plot the average utility of the best-response estimation strategies for the network administrator in Fig. \ref{fig:estimation_strategy}. We observe the following from Fig. \ref{fig:estimation_strategy}.

 \begin{itemize}
   \item When the gain $\admingain$ is low compared to the cost $\admincost$, the network administrator always chooses $\admind$ to be 0 to minimize the cost. As a result, the infection source gets caught only when $\sourced = 0$.
   \item When the gain $\admingain$ is high compared to the cost $\admincost$, the network administrator always chooses $\admind$ to be $\sourced$. As a result, the overall cost increases with $\sourced$ as more nodes need to be investigated, which in turn decreases the utility of the network administrator. Moreover, the infection source always gets caught in this case.
   \item When the gain $\admingain$ is medium compared to the cost $\admincost$, the network administrator chooses $\admind$ to be $\sourced$ when $\sourced$ is small and the gain of identifying the infection source is higher than the cost of investigating more nodes.  When $\sourced$ increases to a point that the increase in cost of investigating $|\Vsp(\admind)|-1$ more nodes exceeds the gain of identifying the infection source, the network administrator chooses $\admind$ to be 0.
 \end{itemize}

\subsection{Incomplete Observations}\label{subsec:incomplete}
In this paper, we have assumed that the network administrator can observe all the infected nodes. In this subsection, we evaluate the robustness of the Jordan center based estimation strategy and DIS infection strategy when only a subset of the infected nodes are observed by the network administrator.

Let $\alpha$ be the percentage of infected nodes that are randomly observed by the network administrator, and let $\JC(\alpha)$ be the Jordan center of the set of observed infected nodes. Note that $\JC(\alpha)$ may be different from the Jordan center of all infected nodes $\JC$. Therefore, the distance $d(v^*, \JC(\alpha))$ can differ from $\sourced$. The network administrator can identify the infection source when $d(v^*, \JC(\alpha)) \le \admind$.

\begin{figure}[!t]
    \centering
    \psfrag{n}[][][0.8][0]{$\alpha = 1$}
    \psfrag{u}[][][0.8][0]{$\alpha = 10$}
    \psfrag{x}[][][0.8][0]{$\alpha = 50$}
    \psfrag{a}[l][][1][0]{$\admind$}
    \psfrag{s}[][][1][0]{$\sourced$}
    \includegraphics[width=1\textwidth]{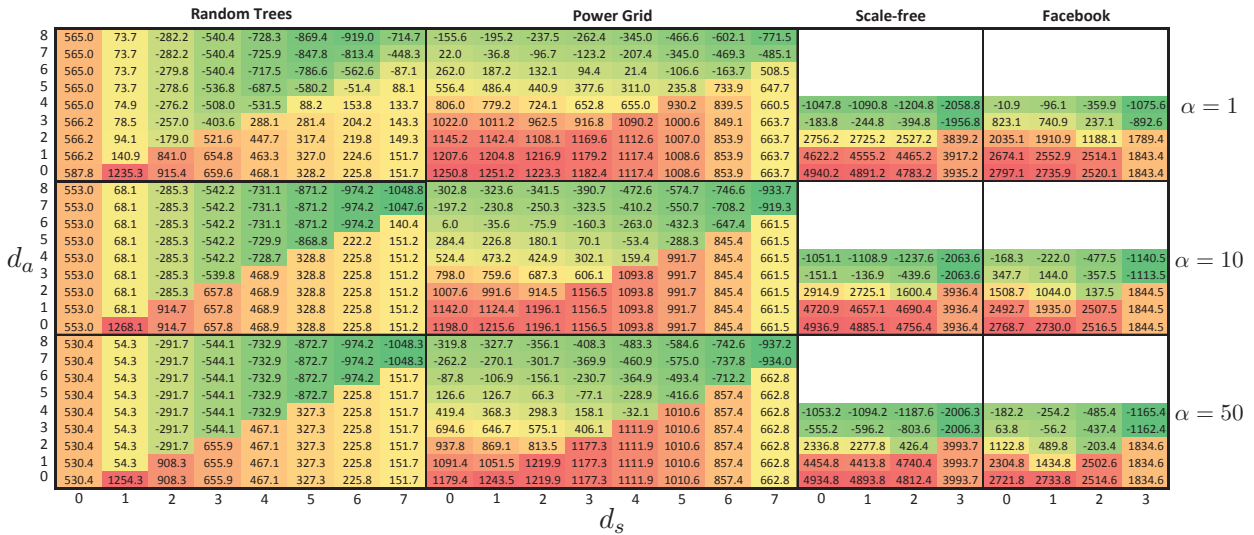}
    \caption{Average utility of the infection source for various observation percentage $\alpha$. The color scale of the heat map is calibrated for each kind of network respectively.}
    \label{fig:incomplete_heatmap_infection_strategy}
\end{figure}

\begin{figure}[!t]
    \centering
    \psfrag{n}[][][0.8][0]{$\alpha = 1$}
    \psfrag{u}[][][0.8][0]{$\alpha = 10$}
    \psfrag{x}[][][0.8][0]{$\alpha = 50$}
    \psfrag{a}[l][][1][0]{$\admind$}
    \psfrag{s}[][][1][0]{$\sourced$}
    \includegraphics[width=1\textwidth]{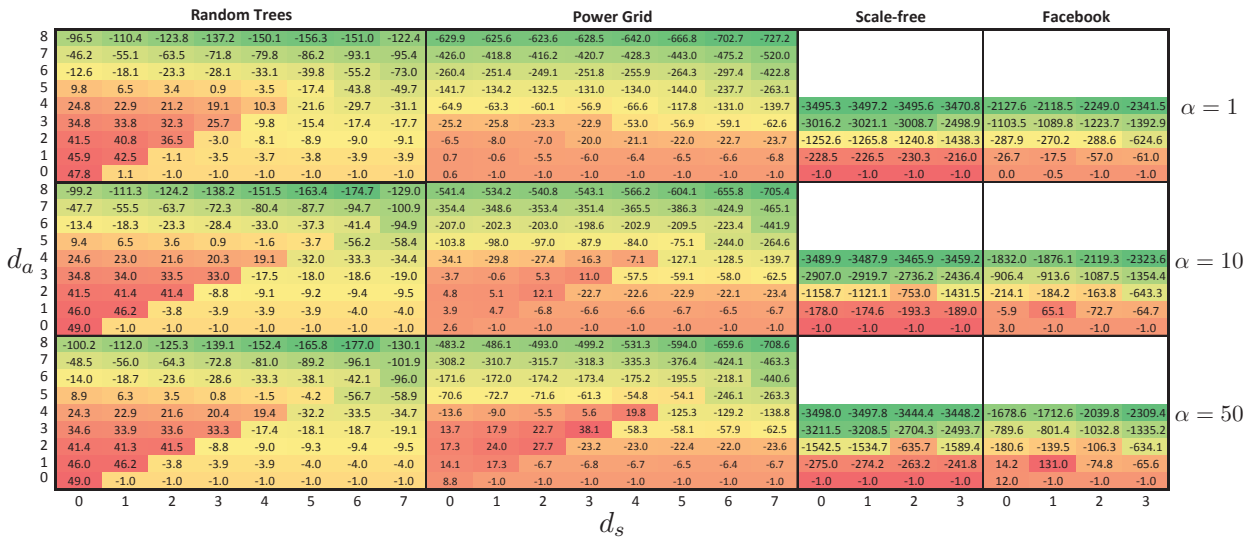}
    \caption{Average utility of the network administrator for various observation percentage $\alpha$. The color scale of the heat map is calibrated for each kind of network respectively.}
    \label{fig:incomplete_heatmap_estimation_strategy}
\end{figure}

We perform simulations on random trees, scale-free networks, the power grid network and the Facebook network. We set $\sourcegain = 1$ and $\sourcecost$ to be 1200, 6000, 1600 and 3000 for random trees, scale-free networks, the power grid network and the Facebook network, respectively. For the network administrator, we let the gain $\admingain$ to be medium compared to $\admincost$. Specifically, we set $\admincost = 1$ and $\admingain$ to be 50, 1500, 200 and 500 for random trees, scale-free networks, the power grid network and the Facebook network, respectively. The observation threshold $\nobs$ for each network is chosen to correspond to the same observation time $\tobs$ used in Section \ref{subsec:number_of_infected}. For each kind of network, we run 1000 simulations for each value of $\sourced \in [0, \bar{\sourced}]$, $\admind \in \{0, 1, \cdots, \sourced+1 \}$ and $\alpha \in \{1, 10, 50\}$, where $\bar{\sourced} = \tobs/2$. For each simulation run, we randomly pick $\alpha$ percent of infected nodes as observed nodes, compute the \emph{realized} utility values of the network administrator and infection source, and average them over the simulation runs. In the realized utilities we compute, the network administrator obtains a gain $g_a$ while the infection source incurs a cost $c_s$, only when $d(v^*, \JC(\alpha)) \le \admind$. The average utilities of the infection source and the network administrator are shown in Fig.~\ref{fig:incomplete_heatmap_infection_strategy} and Fig.~\ref{fig:incomplete_heatmap_estimation_strategy}, respectively.

Consider the utility of the infection source in Fig.~\ref{fig:incomplete_heatmap_infection_strategy}. The best-response of the infection source is still choosing $\sourced$ to be either 0 or $\admind+1$. This implies that the result of Theorem~\ref{theorem:optimal_infection_strategy} is robust for the tested networks even though only a subset of infected nodes can be observed.

Fig.~\ref{fig:incomplete_heatmap_estimation_strategy} shows the utility of the network administrator. For random trees, the best-response of the network administrator is still choosing $\admind$ to be either 0 or $\sourced$, which verifies Theorem~\ref{theorem:optimal_estimation_strategy} in the case where only partial observations are available.
On the other hand, for general networks with incomplete observations of the set of infected nodes, it becomes more difficult for the network administrator to correctly identify the infection source. The network administrator needs to increase $\admind$ in order to have a higher chance of identifying the infection source. However, for dense networks, the cost of probing more nodes can increase very quickly as $\admind$ increases. As a result, for scale-free networks and the Facebook network, the network administrator tends to choose $\admind$ to be 0 to minimize the cost instead. In practice when the network administrator cannot observe all node status, in order to reduce its probing cost, it needs to formulate an estimation strategy that incorporates other side information. For example, in trying to identify the source of a computer virus, part of the cost of examining every node in the suspect set can be reduced by only examining known weak points in the network or by performing a forensic analysis of the virus code to reduce the suspect set size.

\section{Conclusion}\label{sec:conclusion}
We have formulated the problems of maximizing infection spreading and source identification in a network as a strategic game. Conditioned on the strategy of the other player, we proposed best-response strategies for both the infection source and the network administrator in a tree network. We also derived conditions under which a Nash equilibrium exists. In all Nash equilibria, the Jordan center estimator is the equilibrium estimation strategy for the network administrator. We showed that the sum utility of both players is maximized at one of these Nash equilibria.

In this work, we have assumed that the underlying network is a tree. Obtaining theoretical results for general networks seems unlikely due to difficulties in designing an optimal infection strategy in a loopy graph. Future work includes designing best-response strategies for the network administrator under a more general class of estimation strategies that may not be based on the Jordan center. It would also be of interest to study the best-response infection and estimation strategies when infection rates are stochastic and not fully controllable by the infection source, or when additional side information is available to the network administrator. %
We have also adopted a simple game theoretic formulation in this paper where the network administrator makes a one-shot observation of the network. It would be of interest to consider cases where the network administrator can observe the evolution of the network \cite{Jiang2014a,Jiang2014b} by formulating a multi-stage game.

\appendices

\section{Proof of Theorem \ref{theorem:safety_margin_upper_bound}}\label{appendix:theorem:safety_margin_upper_bound}
We prove Theorem \ref{theorem:safety_margin_upper_bound} in two steps. We first show that there exists at least one infection strategy that can achieve $\sourced = \bar{\sourced}$. We then show that there is no infection strategy that results in $\sourced > \bar{\sourced}$.

\textbf{Step 1}: We only need to find one infection strategy that has safety margin $\sourced = \bar{\sourced}$. Let $u \in V$ be a node with $d(v^*,u)=\dt$, and let $D$ be the path from $v^*$ to $u$. Consider the following infection strategy: set the infection rate of each edge in $D$ to be the respective maximum infection rate, and set the infection rates of other edges not in $D$ to be 0. We then have $\sourced = \floor{\dt/2} = \bar{\sourced}$.

\textbf{Step 2}: Assume $\sourced > \bar{\sourced}$, i.e., $\sourced \ge \bar{\sourced}+1$. Consider any infection strategy $\Lambda$ and a Jordan center $u$ such that $d(\sss, u) = \sourced$ and let $D=(l_1, \ldots, u, \ldots, l_2)$ be a diameter of $G_t$ containing $u$, where $l_1$ and $l_2$ are leaf nodes, with $d(l_1,\sss) \le d(l_2,\sss)$. We first show that $d(\sss, u) \le d(l_2,u)$.  It can be shown that $d(l_1,u)$ and $d(l_2,u)$ differs in value by at most 1 \cite{Luo2014}. If $d(l_1, u) = d(l_2,u)+1$, consider the neighbouring node $u'$ of $u$ on the path $\rho(u, l_1)$. From \cite{Luo2014}, we obtain that $u'$ is a Jordan center with $d(\sss,u') = \sourced - 1$, a contradiction. Therefore, we have $d(l_1, u) \le d(l_2,u)$. It is easy to see that $d(\sss,u) \le d(l_2,u)$ because otherwise, the path with $\rho(\sss,u)$ concatenated with $\rho(u,l_2)$ has length greater than that of $D$, a contradiction.  We then have
\begin{align*}
d(\sss,l_2) &= d(\sss,u) + d(u, l_2) \\
&\ge 2 d(\sss,u) \\
&\ge 2 (\bar{\sourced}+1) \\
&\ge \dt + 1,
\end{align*}
a contradiction since the infection can travel at most $\dt$ hops in time $t$. This shows that no infection strategy results in $\sourced > \bar{\sourced}$. The proof for Theorem \ref{theorem:safety_margin_upper_bound} is now complete.

\section{Proof of Lemma \ref{lemma:dominant_path}}\label{appendix:lemma:dominant_path}
We call the leaf node $l_d$ of a dominant path a \emph{dominant leaf}. We prove Lemma \ref{lemma:dominant_path} in two steps. We first show that any diameter $D$ of $\GI[t]$ contains at least one dominant leaf. We then show that the infection rate associated with each edge in any dominant path is its upper bound $\bar{\lambda}_m$.

\textbf{Step 1:} Show that any diameter $D$ of $\GI[t]$ contains at least one dominant leaf.

Let $l_1$ and $l_2$ to be the two end nodes of $D$. Let $v_i$ and $v_j$ to be two different neighboring nodes of $\sss$. We consider two possible scenarios: $l_1$ and $l_2$ are in different subtrees $T_{v_i}$ and $T_{v_j}$, respectively; $l_1$ and $l_2$ are in the same subtree $T_{v_i}$.

\textsc{Scenario 1}: $l_1$ and $l_2$ are in different subtrees $T_{v_i}$ and $T_{v_j}$, respectively.

Suppose $D$ does not contain any dominant leaf. Then we can find a dominant path $\rho(\sss, l_d)$ such that $l_d \notin T_{v_i}$ (if $l_d \in T_{v_i}$, we have $l_d \notin T_{v_j}$ and just exchange the notations $i$ and $j$). Consider the path $D'=\{l_1, \cdots, \sss, \cdots, l_d\}$, we have
\begin{align*}
|D'| & = d(\sss, l_1)+d(\sss, l_d)+1 \\
&>d(\sss, l_1)+d(\sss, l_2)+1 \\
&=|D|.
\end{align*}
Thus, we find a path $D'$ that has longer distance than the diameter $D$, a contradiction. So $D$ must contain at least one dominant leaf.

\begin{figure}[!t]
  \centering
  \psfrag{1}[][][0.9][0]{$\sss$}
  \psfrag{2}[][][0.9][0]{$l_2$}
  \psfrag{3}[][][0.9][0]{$l_1$}
  \psfrag{4}[][][0.9][0]{$l_d$}
  \psfrag{5}[][][0.9][0]{$u_x$}
  \psfrag{6}[][][0.9][0]{$u_y$}
  \psfrag{a}[][][1.1][0]{$T_{v_i}$}
  \includegraphics[width=0.45\textwidth]{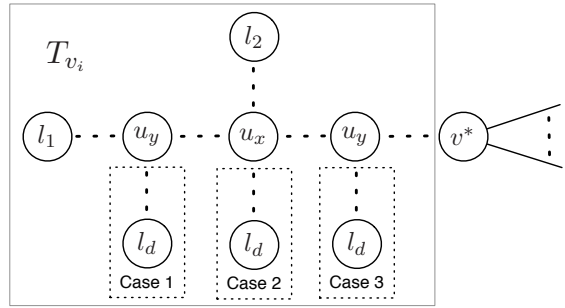}
  \caption{Illustration of part of the infection graph $\GI$.}
\label{fig:DIS_dominant_leaf}
\end{figure}

\textsc{Scenario 2}: $l_1$ and $l_2$ are in the same subtrees $T_{v_i}$.

Suppose $D$ does not contain any dominant leaf. Consider a dominant leaf $l_d$. If $l_d \notin T_{v_i}$, following the same argument as in Scenario 1, we can find a path $D'=\{l_1, \cdots, \sss, \cdots, l_d\}$ that has longer distance than $D$. We now consider the case where $l_d \in T_{v_i}$. Let $u_x$ to be the first node on which the two paths $\rho(l_1, \sss)$ and $\rho(l_2, \sss)$ intersects, where $x$ is the depth of $u_x$ and $1 \le x \le \min \{ d(\sss, l_1)-1, d(\sss, l_2)-1 \}$. Then let $u_y$ to be the first node on which the two paths $\rho(l_1, \sss)$ and $\rho(l_d, \sss)$ intersects, where $y$ is the depth of $u_y$ and $0 \le y \le d(\sss, l_1)-1$. Figure \ref{fig:DIS_dominant_leaf} shows all three possible cases: $y>x, y=x$ and $y<x$. Consider the path $D'=\{l_d, \cdots, u_x, \cdots, l_2\}$. For case 1 and case 2 in Figure \ref{fig:DIS_dominant_leaf}, we have $d(u_x, l_d) > d(u_x, l_1)$ because $d(\sss, l_d) > d(\sss, l_1)$. Similarly, for case 3 in Figure \ref{fig:DIS_dominant_leaf}, we have $d(u_x, l_d)>d(u_y, l_d) > d(u_y, l_1)$. As a result, for all three cases, we have
\begin{align*}
|D'| & = d(u_x, l_d)+d(u_x, l_2)+1 \\
&>d(u_x, l_1)+d(u_x, l_2)+1 \\
&=|D|.
\end{align*}
We find a path $D'$ that has longer distance than the diameter $D$, a contradiction. We can now conclude that any diameter $D$ of $\GI[t]$ contains at least one dominant leaf.

\textbf{Step 2:} Show that the infection rate associated with each edge in any dominant path is its upper bound $\bar{\lambda}_m$.

\begin{figure}[!t]
  \centering
  \psfrag{1}[][][0.9][0]{$\sss$}
  \psfrag{2}[][][0.9][0]{$l_2$}
  \psfrag{3}[][][0.9][0]{$l_d$}
  \psfrag{4}[][][0.9][0]{$u_x$}
  \includegraphics[width=0.28\textwidth]{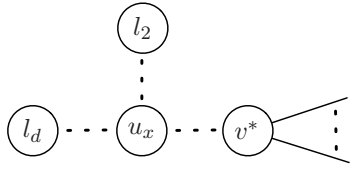}
  \caption{Illustration of the diameter containing the dominant leaf $l_d$.}
\label{fig:DIS_diameter}
\end{figure}

Consider any dominant path $\dpath=\rho(\sss, l_d)$ and suppose the infection rate of some edges in $\dpath$ are less than $\bar{\lambda}_m$. Let $D=\{l_d, \cdots, u_x, \cdots, l_2\}$ to be the diameter containing $l_d$ as shown in Figure \ref{fig:DIS_diameter}, where $x \ge 0$ is the depth of $u_x$. Consider a Jordan center $u$ such that $d(\sss, u) = \sourced$. It is easy to see that $u$ is at the middle of the diameter. Since $d(\sss, l_d) > d(\sss, l_2)$, we have $d(u_x, l_d) > d(u_x, l_2)$, which in turn implies that $u \in \rho(u_x, l_d)$. If we increase the infection rates of all edges in $\dpath$ to their maximum rates, the length of $\dpath$ will increase and $u$ will move further away from $\sss$, i.e., the safety margin will increase as well. We can then increase the infection rates of some edges in the path $\rho(u_x, l_2)$ to increase the length of $\rho(u_x, l_2)$. As a result, $u$ will move closer to $\sss$ and the safety margin can reduce back to its original value. In this case, we find another infection strategy that results in more infected nodes subject to the same safety margin $\sourced$, a contradiction. We can now conclude that the infection rate associated with each edge in any dominant path is its maximum rate. This completes the proof of Lemma \ref{lemma:dominant_path}.

\section{Proof of Lemma \ref{lemma:d1>d2}}\label{appendix:lemma:d1>d2}
Fix any diameter and let it be $D$. We prove Lemma \ref{lemma:d1>d2} by contradiction. Suppose neither end of $D$ is in the subtree $T_{v_1}$, then there are two possible cases: (1) both ends of $D$ are in the subtree $T_{v_i}$, where $2 \leq i \leq k$; (2) the two ends of $D$ are in the subtree $T_{v_i}$ and $T_{v_j}$, respectively, where $2 \leq i,j \leq k$ and $i \ne j$.

We first consider case (1) and let $l_1$ and $l_2$ to be the two ends of $D$. Consider two paths $\rho(l_1,\sss)$ and $\rho(l_2,\sss)$ and let the first node on which these two paths intersect as $w$. We can find a path $D'=\{\leaf{v_1}, \cdots, \sss, \cdots, w, \cdots, l_2\}$ that has longer distance than $D$. We have
\begin{align*}
&|\{ \leaf{v_1}, \cdots, \sss, \cdots, w\}| \\
&> |\{ \leaf{v_1}, \cdots, \sss\}| \\
&= d(\sss,\leaf{v_1})+1 \\
&>d(\sss,\leaf{v_i})+1 \\
&>d(l_1, u)+1 \\
&=|\{ l_1,\cdots,u \}|.
\end{align*}
Then we have
\begin{align*}
|D'| &= | \{ \leaf{v_1}, \cdots, \sss, \cdots, u \} | + d(u,l_2) \\
&> | \{ l_1,\cdots,u \} |+d(u,l_2) \\
&= |D|.
\end{align*}
We find a path that has greater length than the diameter, which contradicts with the definition of diameter. This completes the proof for case (1).

We then consider case (2). Fix $i$ and $j$, and let $l_i$ and $l_j$ denote the two ends of the diameter in subtree $T_{v_i}$ and $T_{v_j}$, respectively. Since $d(\sss,\leaf{v_1}) > d(\sss,\leaf{v_i})$, the length of the path $D'=\{\leaf{v_1}, \cdots, \sss, \cdots, \leaf{v_j}\}$ is greater than the length of diameter $D=\{\leaf{v_i}, \cdots, \sss, \cdots, \leaf{v_j}\}$, which contradicts with the definition of diameter. This completes the proof for case (2), and the proof for Lemma \ref{lemma:d1>d2} is now complete.

\section{Proof of Lemma \ref{lemma:DIS_Pd}}\label{appendix:lemma:DIS_Pd}
Consider any infection strategy $\Lambda$ with safety margin $d_s$, a dominant path $\dpath$, and that maximizes the number of infected nodes. We use the same notations in the discussion preceding Lemma \ref{lemma:DIS_Pd}, with $\dpath = (u_0,\ldots,u_{\dt})$. In addition, let $T_u(G)$ be the subtree of $G$ rooted at node $u$ with the first link in the path from $u$ to $v^*$ removed. Let $\tilde{T}_{u_m}(G)$ be $T_{u_m}(G)\backslash T_{u_{m+1}}(G)$.

Suppose there exists a $m \in [0,d_s]$ such that for a path $\rho(u_m,\tleaf{u_m})$, we have $d(u_m , T_{u_m}(G)) \geq \lambda_m'(t-t_m)$, and $\lambda_m' > \lambda_m$, where
\begin{align*}
\lambda_m' = d(u_m,\tleaf{u_m})\left( \sum_{(i,j)\in\rho(u_m,\tleaf{u_m})} \lambda(i,j)^{-1} \right)^{-1},
\end{align*}
i.e., $\lambda_m'$ is the average infection rate along the path $\rho(u_m,\tleaf{u_m})$. (Recall that $\lambda_m$ is the average infection rate used by $\DIS[d_s,t]$ for $\tilde{T}_{u_m}(G)$.) We have
\begin{align*}
\left\lfloor \lambda_m' (t-t_m)\right\rfloor = d(u_m,\tleaf{u_m}),
\end{align*}
since otherwise, we can infect more nodes, contradicting the assumption that $\Lambda$ maximizes the number of infected nodes. We also have $\est(u_m)$ is a Jordan center.
By replacing $\lambda_m$ with $\lambda_m'$ in \eqref{eqn:u_m_safety_margin}, we have $d(\sss, \est(u_m)) < d_s$ since $\lambda_m' > \lambda_m$ implies that the inequality in \eqref{eqn:lambda_m_upper_bound} is reversed when $\lambda_m$ is replaced by $\lambda_m'$. This contradicts the assumption that $\Lambda$ has safety margin $d_s$.

On the other hand, if $d(u_m , \tilde{T}_{u_m}(G)) < \lambda_m'(t-t_m)$, i.e., all the nodes in $\ttree{u_m}$ are infected by $\Lambda$ before time $t$, then we can choose a $\lambda_m'' \in (\lambda_m', \lambda_m)$, infect all the nodes in $\ttree{u_m}$ by time $t$, and repeat the above argument using $\lambda_m''$ in place of $\lambda_m'$. Therefore, no other strategy can infect more nodes than $\DIS[d_s,t]$, and the proof is complete.

\section{Proof of Theorem \ref{theorem:nash_equilibrium}}\label{appendix:theorem:nash_equilibrium}

We first prove the properties \eqref{theorem:nash_equilibrium:00}-\eqref{theorem:nash_equilibrium:no} in sequence. We then prove the sum utility optimality claim.

\noindent \textbf{Proof of Theorem \ref{theorem:nash_equilibrium}\eqref{theorem:nash_equilibrium:00}.}

Following the definition of the Nash equilibrium, it suffices to show that
\begin{align}
u_a(0,\MIS{0}) \ge u_a(\admind, \MIS{0}), \ &\forall  \admind > 0, \label{eqn:nash_equilibrium:00:ua} \\
u_s(0,\MIS{0}) \ge u_s(0, \Lambda), \ &\forall  \Lambda. \label{eqn:nash_equilibrium:00:us}
\end{align}

The inequality \eqref{eqn:nash_equilibrium:00:ua} follows from Theorem \ref{theorem:optimal_estimation_strategy}. To show \eqref{eqn:nash_equilibrium:00:us}, let $\Lambda$ be any infection strategy, and $\sourced$ be its safety margin. If $\sourced = 0$, we obtain from Lemma \ref{lemma:maximum_infection_strategy_relationship}
\begin{align*}
u_s(0,\MIS{0}) &=\sourcegain |\VI(\MIS{0})| - \sourcecost(0)\\
&\ge \sourcegain |\VI(\Lambda)| - \sourcecost(0)\\
&=u_s(0, \Lambda).
\end{align*}
If $\sourced > 0$, from the assumption of Theorem \ref{theorem:nash_equilibrium}\eqref{theorem:nash_equilibrium:00} and Lemma \ref{lemma:maximum_infection_strategy_relationship}, we have for any $\Lambda_1\in \MISSET{1}$,
\begin{align*}
u_s(0,\MIS{0}) &\ge u_s(0, \MIS{1}) \\
&=\sourcegain |\VI(\MIS{1})|\\
&\ge \sourcegain |\VI(\Lambda)|\\
&=u_s(0, \Lambda).
\end{align*}

The proof of Theorem \ref{theorem:nash_equilibrium}\eqref{theorem:nash_equilibrium:00} is now complete.

\noindent \textbf{Proof of Theorem \ref{theorem:nash_equilibrium}\eqref{theorem:nash_equilibrium:01}.}

It again suffices to show that for $\Lambda_1\in \MISSET{1,\tobs}$ satisfying the assumptions of Theorem \ref{theorem:nash_equilibrium}\eqref{theorem:nash_equilibrium:01}, we have
\begin{align}
u_a(0,\MIS{1}) \ge u_a(\admind, \MIS{1}), \ &\forall  \admind > 0, \label{eqn:nash_equilibrium:01:ua} \\
u_s(0,\MIS{1}) \ge u_s(0, \Lambda), \ &\forall  \Lambda. \label{eqn:nash_equilibrium:01:us}
\end{align}

From the second assumption of Theorem \ref{theorem:nash_equilibrium}\eqref{theorem:nash_equilibrium:01}, for any $\admind > 0$, we have
\begin{align*}
u_a(0,\MIS{1}) & \ge u_a(1,\MIS{1}) \\
& = -\admincost(\Vsp(1)) + \admingain(1,\VI(\MIS{1})) \\
& \ge -\admincost(\Vsp(\admind)) + \admingain(\admind,\VI(\MIS{1})) \\
& = u_a(\admind,\MIS{1}),
\end{align*}
and inequality \eqref{eqn:nash_equilibrium:01:ua} holds.

We now show \eqref{eqn:nash_equilibrium:01:us}. Let $\Lambda$ be any infection strategy and $\sourced$ be its safety margin. If $\sourced = 0$, from the first assumption of Theorem \ref{theorem:nash_equilibrium}\eqref{theorem:nash_equilibrium:01}, and Definition \ref{def:maximum_infection_strategy}, we have
\begin{align*}
u_s(0,\MIS{1}) &\ge u_s(0, \MIS{0}) \\
&=\sourcegain |\VI(\MIS{0})| - \sourcecost(0)\\
&\ge \sourcegain |\VI(\Lambda)| - \sourcecost(0)\\
&=u_s(0, \Lambda).
\end{align*}

If $\sourced > 0$, from Lemma \ref{lemma:maximum_infection_strategy_relationship}, we have
\begin{align*}
u_s(0, \MIS{1}) &=\sourcegain |\VI(\MIS{1})|\\
&\ge \sourcegain |\VI(\Lambda)|\\
&=u_s(0, \Lambda).
\end{align*}
We have now shown that \eqref{eqn:nash_equilibrium:01:us} holds and the proof of Theorem \ref{theorem:nash_equilibrium}\eqref{theorem:nash_equilibrium:01} is complete.

\noindent \textbf{Proof of Theorem \ref{theorem:nash_equilibrium}\eqref{theorem:nash_equilibrium:no}.}

Consider any strategy pair $(\admind, \Lambda) \ne (0,\Lambda_0)$ or $(0,\Lambda_1)$ for all $\Lambda_1\in \MISSET{1,\tobs}$. Let $\sourced$ be the safety margin of $\Lambda$. If $d_a=0$ and $d_s=0$, then
\begin{align*}
u_s(0,\Lambda)&= g_s|\VI(\Lambda)| - c_s(0)\\
& < g_s|\VI(\Lambda_0)| - c_s(0) \\
& = u_s(0,\Lambda_0),
\end{align*}
so $(\admind, \Lambda)$ is not a Nash equilibrium. If $d_a=0$ and $d_s \geq 1$, we have $u_s(0,\Lambda) = g_s|\VI(\Lambda)| < g_s|\VI(\Lambda_1)|$ for any $\Lambda_1\in \MISSET{1,\tobs}$. This again implies that $(\admind, \Lambda)$ is not a Nash equilibrium.

Now suppose that $d_a > 0$. From Theorem \ref{theorem:optimal_infection_strategy}, it suffices to show that $(d_a,\Lambda_{0})$ and $(d_a,\Lambda_{d_a+1})$ for all $\Lambda_{d_a+1}\in \MISSET{d_a+1,\tobs}$ are not Nash equilibria. We have
\begin{align*}
u_a(d_a,\MIS{0}) &= -\admincost(\Vsp(\admind)) + \admingain(\admind, \VI(\MIS{0}) \\
&< -\admincost(\Vsp(0)) + \admingain(0, \VI(\MIS{0})) \\
& = u_a(0,\MIS{0}),
\end{align*}
where the inequality follows from the assumption that $\bar\sourced > 0$. For each $\Lambda_{d_a+1}\in \MISSET{d_a+1,\tobs}$, we have
\begin{align*}
u_a(d_a,\MIS{d_a+1}) &= -\admincost(\Vsp(\admind+1))\\
&< -\admincost(\Vsp(0))\\
&= u_a(0, \MIS{d_a+1}).
\end{align*}
This completes the proof of Theorem \ref{theorem:nash_equilibrium}\eqref{theorem:nash_equilibrium:no}.

\noindent \textbf{Proof of sum utility optimality.}

It suffices to show that for any strategy pair $(\admind, \Lambda)$ and every $\Lambda_1\in\MISSET{1,\tobs}$, we have
\begin{align}
u_a(\admind, \Lambda)+u_s(\admind, \Lambda) \le
\max \{u_a(0, \MIS{0})+u_s(0, \MIS{0}),\ u_a(0, \MIS{1})+u_s(0, \MIS{1})\}. \label{eqn:sum_utility}
\end{align}

Let $d_s$ be the safety margin of $\Lambda$. We first consider the case where $\admind < \sourced$. Following \eqref{eqn:admin_utility}, \eqref{eqn:source_utility}, Definition \ref{def:maximum_infection_strategy} and Lemma \ref{lemma:maximum_infection_strategy_relationship}, we have for any $\Lambda_1\in \MISSET{1,\tobs}$,
\begin{align*}
u_a(\admind, \Lambda)+u_s(\admind, \Lambda) & = -\admincost(\admind) + \sourcegain |\VI(\Lambda)|  \\
& \le -\admincost(0) + \sourcegain |\VI(\MIS{1})| \\
&= u_a(0, \Lambda_1)+u_s(0, \MIS{1}).
\end{align*}
This implies that \eqref{eqn:sum_utility} holds for all $\Lambda_1\in \MISSET{1,\tobs}$.

Suppose now that $\admind \ge \sourced$. Following \eqref{eqn:admin_utility}, \eqref{eqn:source_utility}, Definition \ref{def:maximum_infection_strategy} and Lemma \ref{lemma:maximum_infection_strategy_relationship}, we have
\begin{align*}
u_a(\admind, \Lambda)+u_s(\admind, \Lambda) & = -\admincost(\admind) +\admingain(\admind) + \sourcegain |\VI(\Lambda)| -\sourcecost(\admind) \\
& \le -\admincost(0) +\admingain(0) + \sourcegain |\VI(\MIS{0})|  -\sourcecost(0)\\
&= u_a(0, \MIS{0})+u_s(0, \MIS{0}).
\end{align*} This implies that \eqref{eqn:sum_utility} holds. The proof of Theorem \ref{theorem:nash_equilibrium} is now complete.

\bibliography{IEEEabrv,SIS}{}
\bibliographystyle{IEEEtran}
\end{document}